\documentclass[%
 aip,
 amsmath,amssymb,
 reprint,%
]{revtex4-2}

\usepackage{graphicx}
\usepackage{dcolumn}
\usepackage{bm}

\usepackage[utf8]{inputenc}
\usepackage[T1]{fontenc}
\usepackage{mathptmx}
\usepackage{etoolbox}
\usepackage{comment}
\usepackage{xcolor}
\makeatletter
\def\@email#1#2{%
 \endgroup
 \patchcmd{\titleblock@produce}
  {\frontmatter@RRAPformat}
  {\frontmatter@RRAPformat{\produce@RRAP{*#1\href{mailto:#2}{#2}}}\frontmatter@RRAPformat}
  {}{}
}%
\makeatother
\begin{document}

\preprint{AIP/123-QED}

\title{Fabrication-tolerant frequency conversion in thin film lithium niobate waveguide with layer-poled modal phase matching}

\author{O. Hefti}
 \altaffiliation[Also at ]{EPFL, Photonic Systems Laboratory (PHOSL), 1015 Lausanne (CH).}
\author{J.-E. Tremblay}%
\author{A. Volpini}
 \email{olivia.hefti@csem.ch}
\affiliation{ 
CSEM SA, Rue de l'Observatoire 58, 2000 Neuchâtel (CH).
}%

\author{Y. Koyaz}
\affiliation{%
EPFL, Photonic Systems Laboratory (PHOSL), Rte Cantonale, 1015 Lausanne (CH).
}%

\author{I. Prieto}
\author{O. Dubochet}
\author{M. Despont}
\author{H. Zarebidaki}
\author{C. Caër}
\author{J. Berney}
\author{S. Lecomte}
\author{H. Sattari}%
\affiliation{ 
CSEM SA, Rue de l'Observatoire 58, 2000 Neuchâtel (CH).
}%

\author{C.-S. Brès}
\affiliation{%
EPFL, Photonic Systems Laboratory (PHOSL), Rte Cantonale, 1015 Lausanne (CH).
}%

\author{D. Grassani}
\affiliation{ 
CSEM SA, Rue de l'Observatoire 58, 2000 Neuchâtel (CH).
}%

\date{\today}

\begin{abstract}
Thanks to its high quadratic nonlinear susceptibilty and low propagation losses, thin film lithium niobate (TFLN) on insulator is an ideal platform for laser frequency conversion and generation of quantum states of light. Frequency conversion is usually achieved by quasi-phase matching (QPM) via electric-field poling. However, this scheme shows very high sensitivity  to the dimensions of the waveguide, poling period and duty cycle, resulting in a lack of repeatability of the phase matched wavelength and efficiency, which in turn limits the spread of TFLN frequency converters in complex circuits and hinders wafer-scale production.
Here we propose a layer-poled modal phase matching (MPM) that is 5 to 10 times more robust towards fabrication uncertainties and theoretically more efficient than conventional QPM. By selectively poling the bottom part of the waveguide all along its length, second harmonic is efficiently generated on a higher order waveguide's mode. We validate this approach by poling TFLN waveguides as a post-process after the fabrication in a foundry process. We perform a tolerance analysis and compare the experimental results with conventional QPM second harmonic generation process on the same waveguides. Then, we show how MPM can be exploited to obtain efficient intraband frequency conversion processes at telecom wavelengths by leveraging simultaneous second harmonic and difference frequency generation in the same waveguide.

\end{abstract}
\maketitle

\section{\label{sec:level1}Introduction:\protect\\ }
Nonlinear optical frequency conversion plays a crucial role in generating light at wavelengths that conventional gain materials can not access \cite{boes2023lithium}. It is also essential for all-optical signal processing \cite{willner2013all} and for the creation of quantum states of light \cite{xin2022spectrally}. Second-order nonlinear platforms are particularly suitable for bridging distant spectral regions and they exhibit intrinsically higher efficiency with respect to third-order nonlinear materials.

Recently, thin film lithium niobate (TFLN) has emerged as a promising platform for photonic integrated circuits (PICs), showing high second-order ($\chi^{(2)}$) nonlinear operations from visible to mid-IR \cite{Ludwig2023ultraviolet}, and enabling seamless integration with other optical components at chip scale \cite{stokowski2023integrated}.

\begin{figure*}[ht]
\centering
    \begin{tabular}{ccc}
     \includegraphics[width=0.22\linewidth]{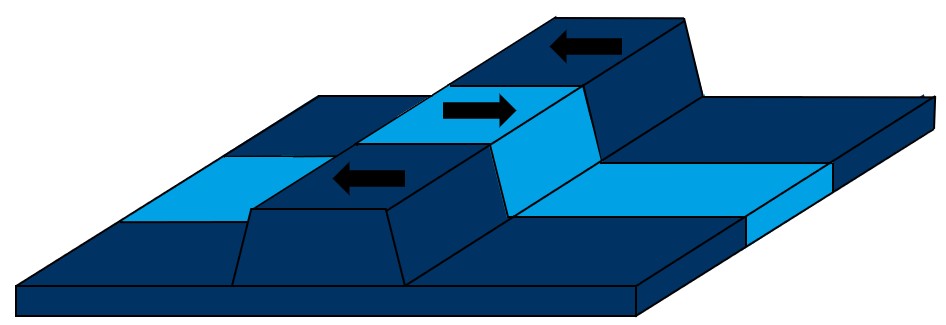} &
       \includegraphics[width=0.22\linewidth]{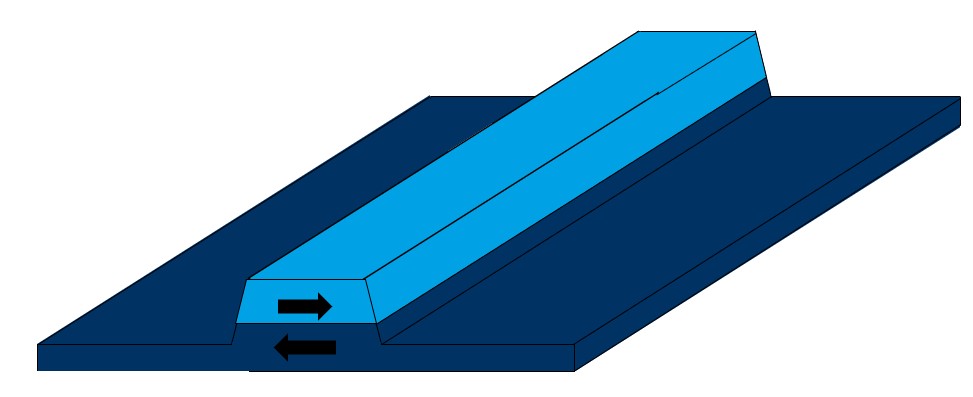} &
        \includegraphics[width=0.43\linewidth]{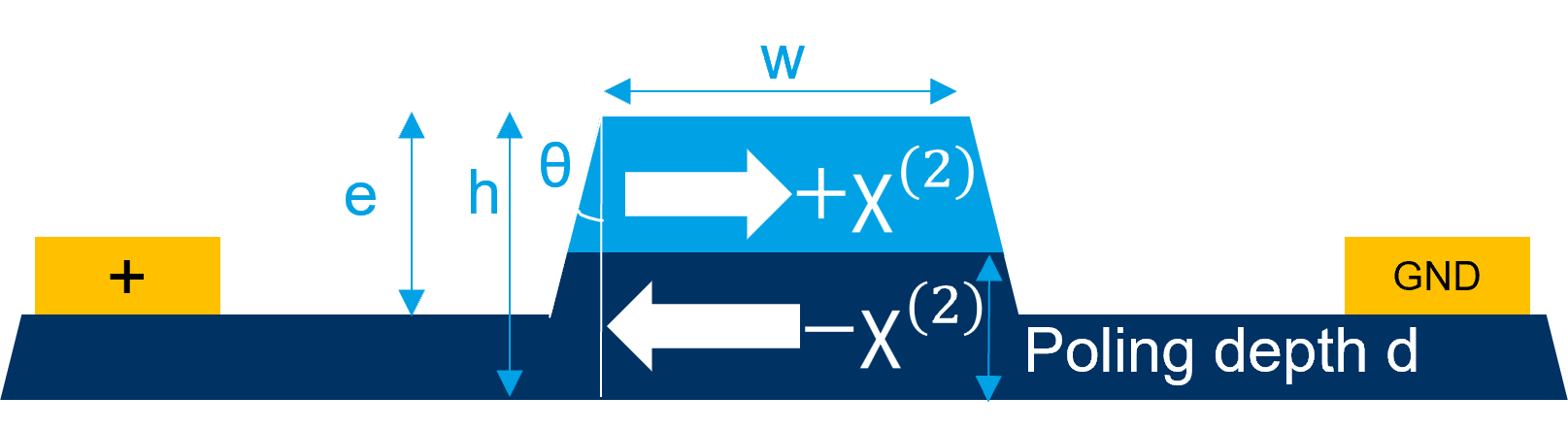}
         \\
        \textbf{(a)} & \textbf{(b)} & \textbf{(c)}  \\[6pt]
    \end{tabular}
    
    \begin{tabular}{cc}
 \includegraphics[width=0.45\linewidth,height=\textheight,keepaspectratio]{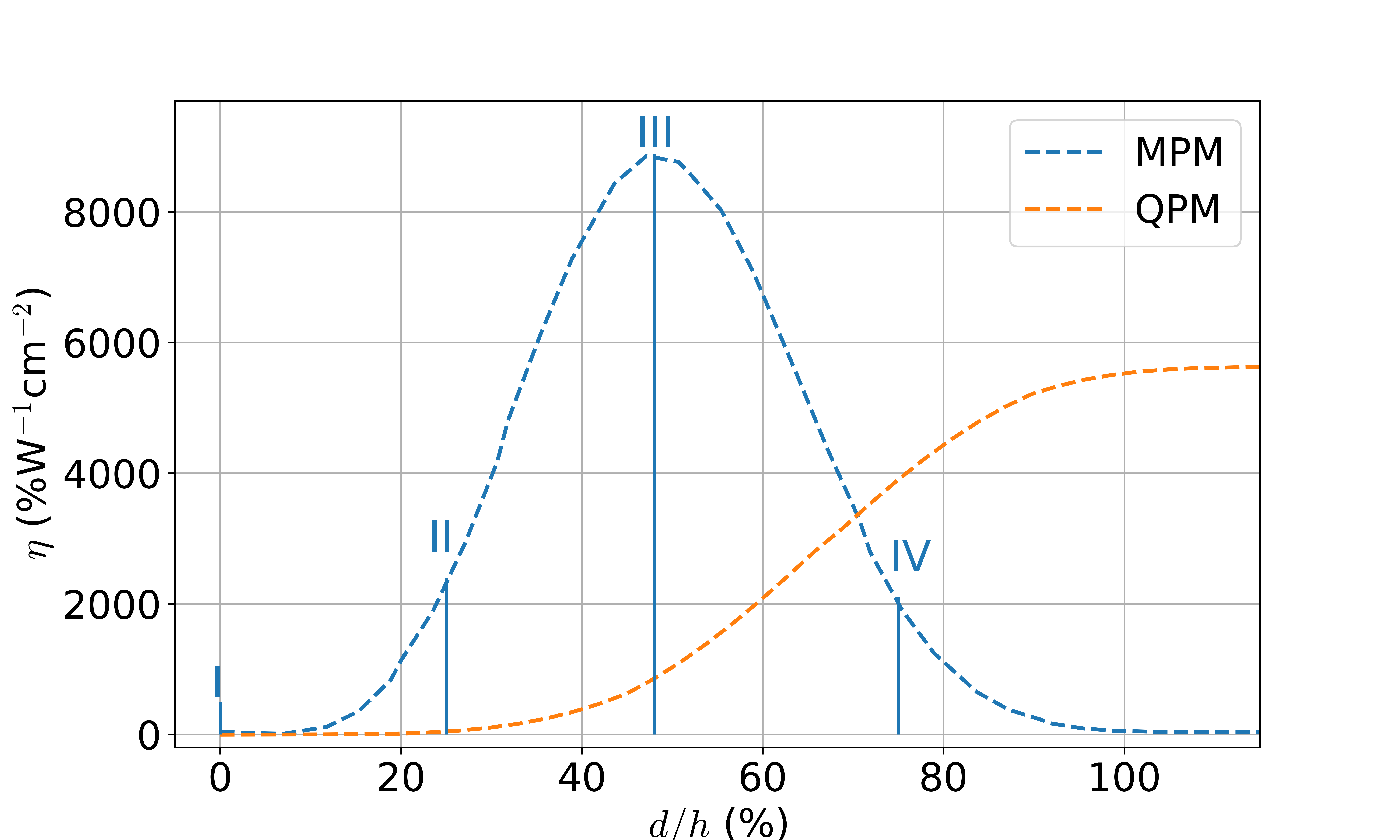} &
    \includegraphics[width=0.45\linewidth]{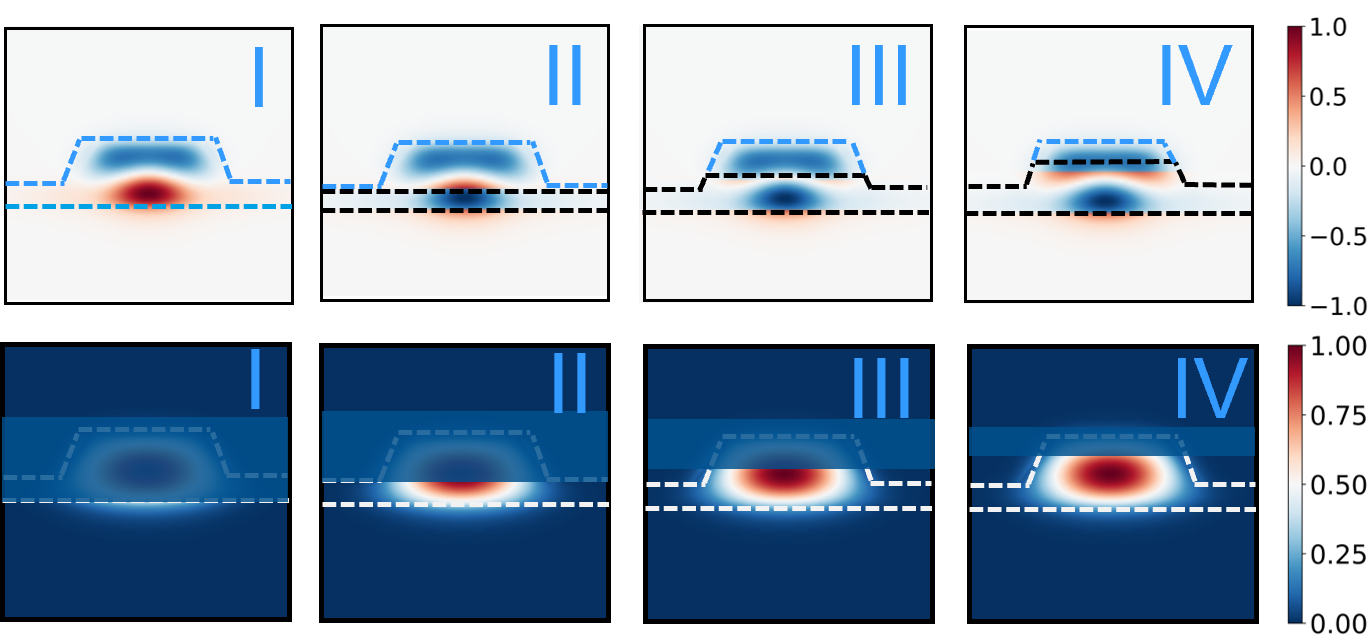}  
       \\
          \textbf{(d)} & \textbf{(e)} \\[6pt]
    \end{tabular}
   
    \caption
      {  Poling configuration maximizing the SH conversion efficiency in the case of \textbf{(a)} QPM between fundamentals and \textbf{(b)} MPM with the TE$_{01}$ mode. \textbf{(c)} Cross-section of the waveguide showing the relevant parameters for QPM and MPM. \textbf{(d)} Simulated SHG conversion efficiency as a function of the poling depth for MPM and QPM.  \textbf{(e)} Real part of the x-component of the electric fields at different poling depths (black dotted lines) for the TE$_{01}$ mode (top) and section of the SH TE$_{00}$ mode contributing to QPM (unblurred) at different poling depths (bottom).  }
   
    \label{fig:birds}
\end{figure*}

The efficiency of nonlinear frequency conversion processes relies on ensuring both energy conservation and phase matching of the interacting waves, to maintain coherent nonlinear interaction. While energy conservation is limited by the transparency window of the material, phase matching between waves with high frequency separation is prevented by the strong frequency dependence of the waveguide effective index induced by dispersion. Quasi-phase matching (QPM) allows frequency conversion on the fundamental mode by compensating the missing momentum through a periodic reversal of the second-order nonlinear polarization of the waveguide \cite{boyd2008nonlinear}. In the case of TFLN, QPM is normally achieved by electric-field poling, which consists in applying strong electrical pulses through patterned electrodes to periodically reverse the polarity of the ferroelectric domains along the waveguide's propagation direction, as illustrated in Fig. \ref{fig:birds} a).

Although there have been many demonstrations of QPM in TFLN \cite{zhu2021integrated,zhao2020shallow,rao2019actively,koyaz2024ultrabroadband,wang2018ultrahigh,jankowski2021dispersion}, achieving a robust and reproducible periodic poling is still a major challenge \cite{xin2024wavelength}. First, the conversion efficiency depends on the poling quality, i.e. a precise duty cycle and a full inversion of the domains, along the height and width of the waveguide, in the poled regions. Second, the QPM wavelength and associated poling period is extremely sensitive to the waveguide dimensions \cite{boes2021efficient}. Discrepancies between the designed and actual waveguide geometry can dramatically shift the QPM wavelength. Reproducibility of the phase matching condition becomes essential as the circuit complexity grows, where large sweeps of the waveguide parameters or of the poling period become unrealistic. Performing electric-field poling after etching the waveguides has been suggested to adapt the poling period accordingly to the measured cross-section \cite{xin2024wavelength}. Although effective in improving the accuracy of the phase matched wavelength, poling etched waveguides leads to poor electric field penetration in the ridge part and, in turn, lowers the conversion efficiency. This phenomenon especially affects deeply etched waveguides, with a large ridge to slab thickness ratio. Recently, QPM has been achieved with heterogeneous waveguides where PPLN films are deposited by micro-transfer printing on top of SiN waveguides. Access to the planar interface allows to blue (red)-shift the QPM wavelength by cladding deposition (thinning the printed PPLN) \cite{vandekerckhove2025scalable}.

An alternative approach relies on phase matching among the fundamental and high order modes of the waveguide, a process known as modal phase matching (MPM). MPM usually results in lower efficiency, due to the poor mode field overlap between the pump and the second harmonic (SH) mode. However, recently, it has been proposed to introduce a vertical asymmetry in the nonlinear polarization of the material, as illustrated in Fig. \ref{fig:birds} b), to greatly increase the nonlinear overlap among the fundamental quasi-TE$_{00}$ mode of the pump and the quasi-TE$_{01}$ mode of the SH.

Such approach has been firstly proposed by using heterogeneous integration with material lacking second-order nonlinearity \cite{luo2019semi}, or by bonding two TFLN layers with opposite polarizations \cite{zhang2020antisymmetric,wang2021efficient}. The latter showing efficiencies as high as 5540 \% W$^{-1}$cm$^{-2}$ in a single pass waveguide and 440'000 \% W$^{-1}$ in resonators \cite{wu2023second,du2024high}. 
Recently, it has been shown that poling etched waveguides is a viable way to introduce such vertical asymmetry in the TFLN \cite{hefti2024symmetry,shi2024efficient}. Such poling does not require any periodicity but has to be constant along the waveguide length. Importantly, this method does not require any heterogeneous integration or additional fabrication step, does not increase the propagation losses of the waveguide, and can be applied to selected areas at the back end of the fabrication process, without affecting other components on the chip.

Building up on this result, here we theoretically and experimentally validate how this layer-poled MPM approach can provide enhanced robustness to fabrication tolerances for second harmonic generation (SHG) compared to standard QPM, without sacrificing conversion efficiency. Finally, we demonstrate efficient and broadband intraband frequency conversion by leveraging a cascaded processes involving SHG and difference frequency generation (DFG) within the same waveguide by injecting two lasers in the C-band (1530-1565 nm) and generating a third frequency in the same band. The results are obtained in PIC waveguides fabricated in a wafer-level foundry process.

\section{\label{sec:level2}Theory:\protect\\  }
\subsection{\label{sec:level2}Second harmonic generation efficiency}

\begin{figure*}[ht]
\centering
    \begin{tabular}{cc}
    \includegraphics[width=0.43\linewidth,height=\textheight,keepaspectratio]{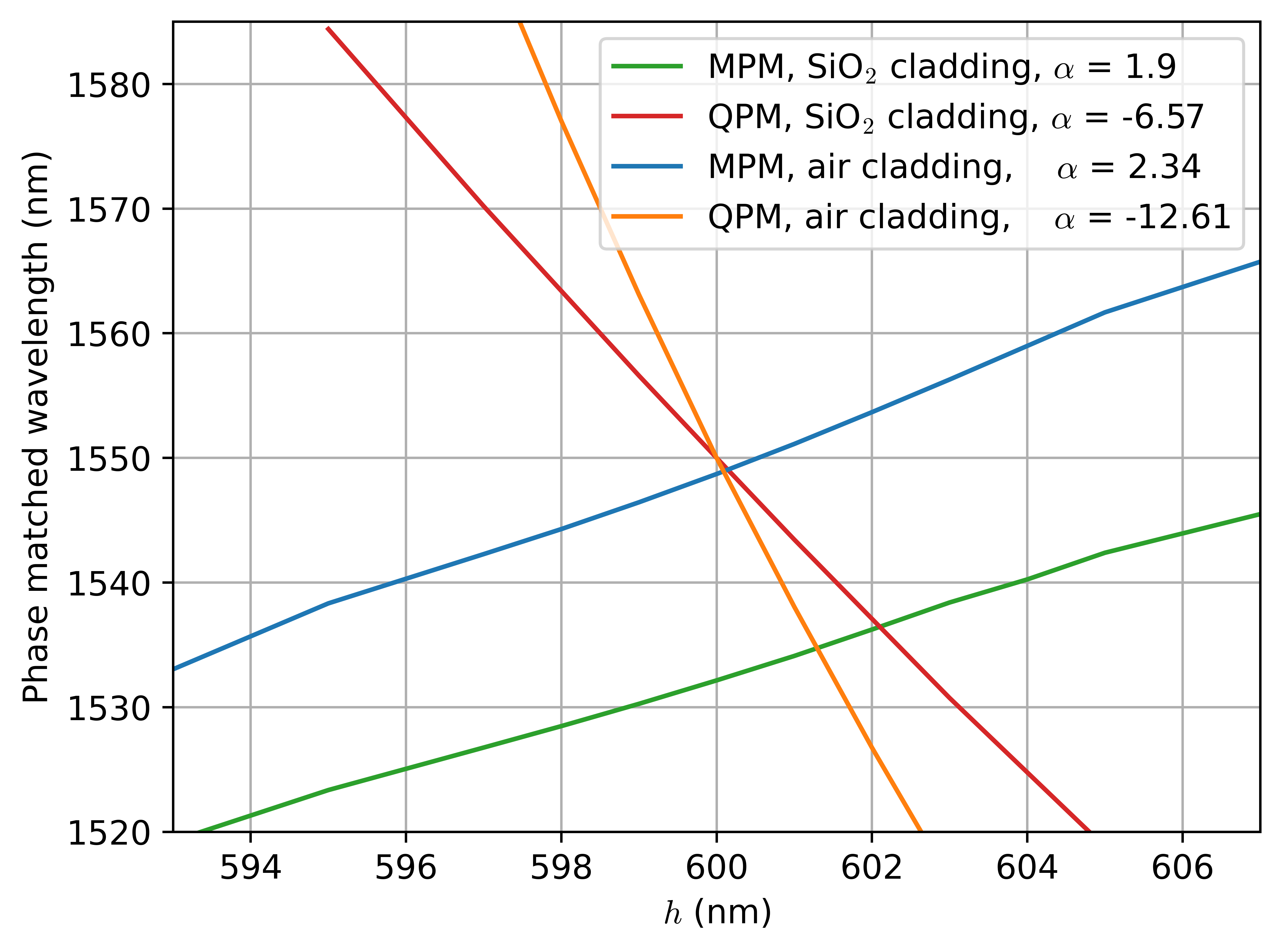} &
    \includegraphics[width=0.445\linewidth,height=\textheight,keepaspectratio]{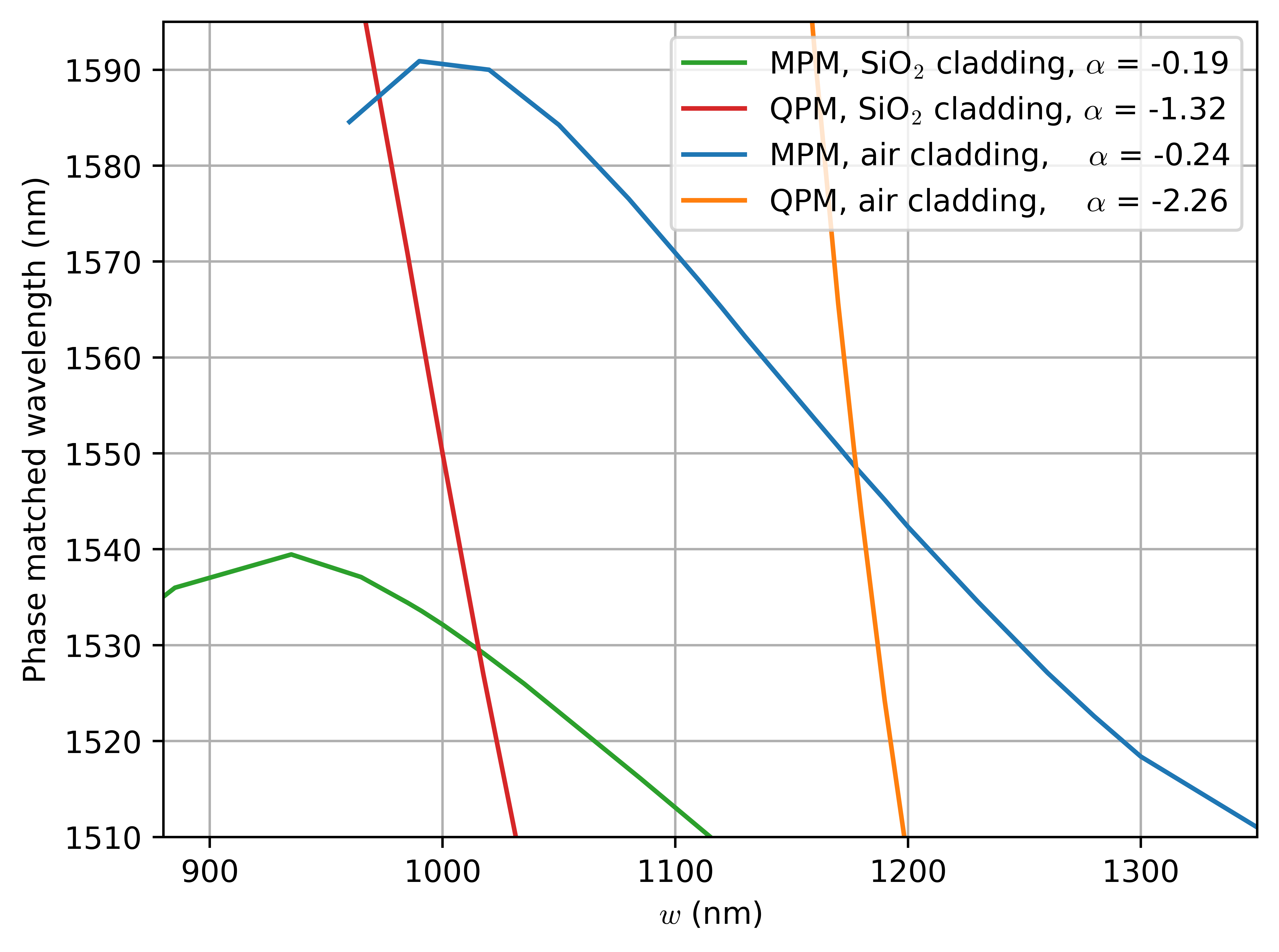} \\
     \\
         \textbf{(a)} & \textbf{(b)}  \\[6pt]
    \end{tabular}

    \begin{tabular}{cc}
    \includegraphics[width=0.47\linewidth,height=\textheight, keepaspectratio]{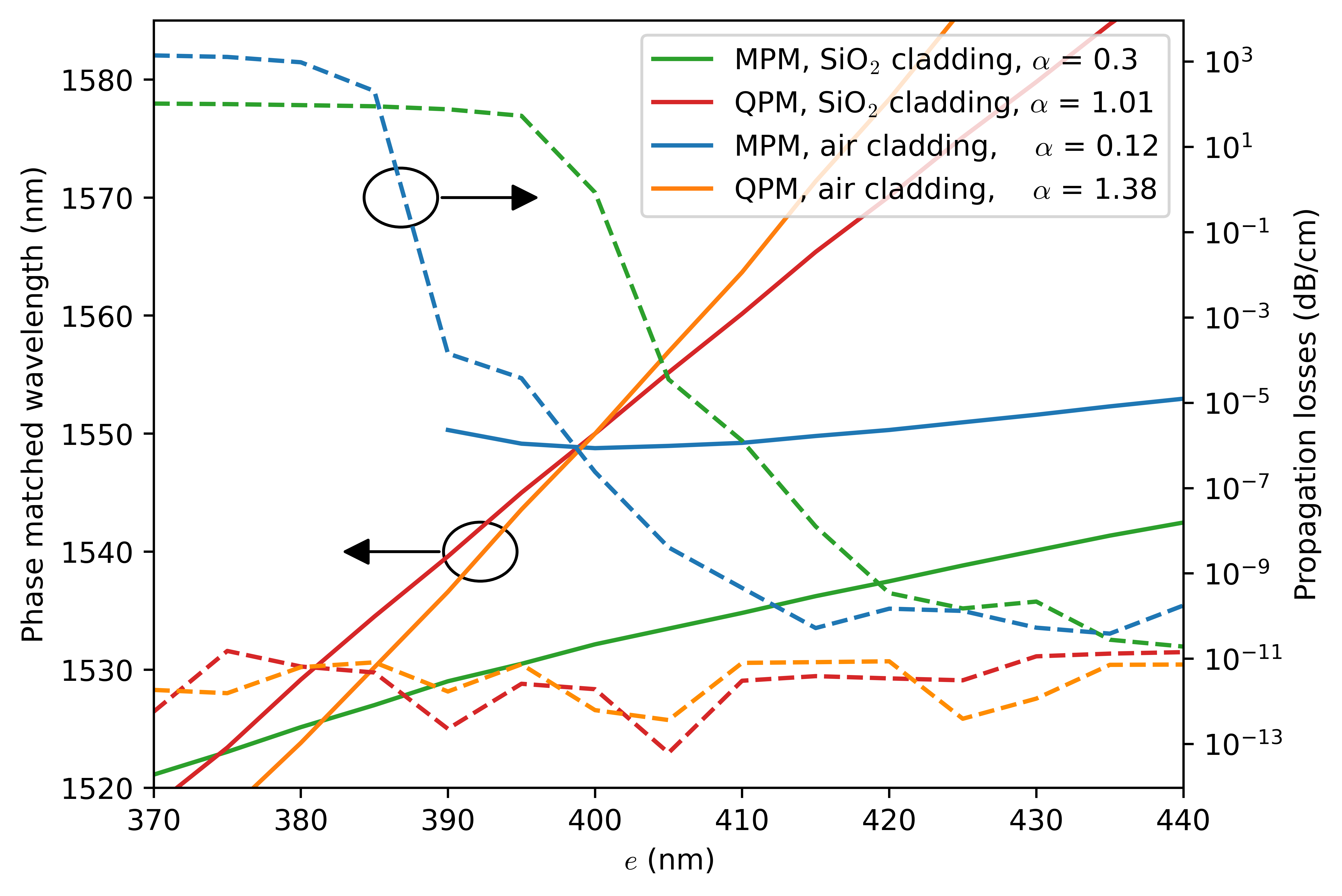}&
      \includegraphics[width=0.43\linewidth,height=\textheight, keepaspectratio]{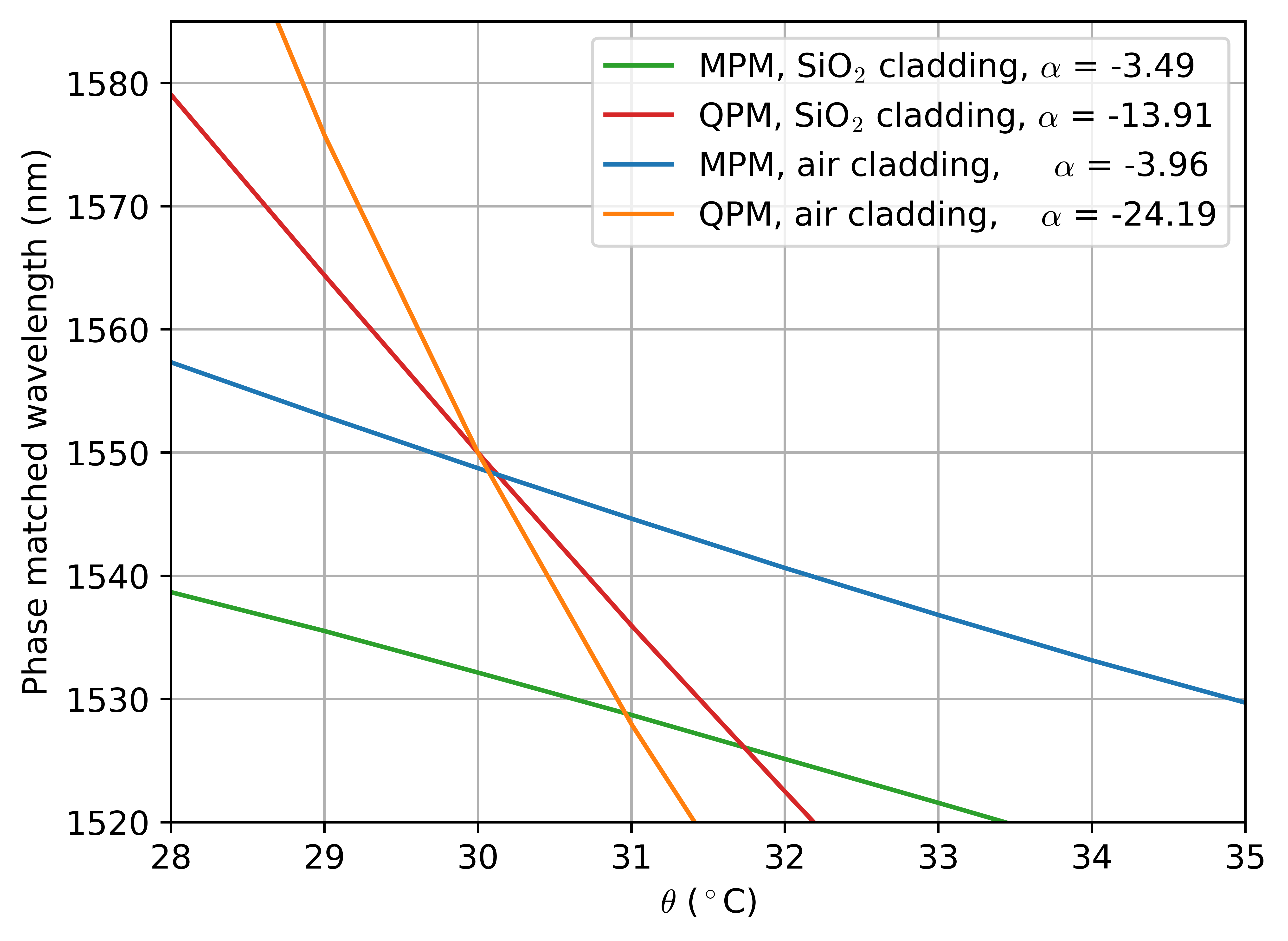} 
     \\
        \textbf{(c)}  & \textbf{(d)} \\[6pt]
    \end{tabular}
    
    \caption
    {Simulated phase matched wavelengths in the case of QPM and MPM as a function of \textbf{(a)} the waveguide height, \textbf{(b)} the waveguide top width, \textbf{(c)} the etch depth (solid lines, left axis) and \textbf{(d)} the fabrication angle. \textbf{(c)} also shows the propagation losses due to leakage in the slab of the TE$_{01}$ and TE$_{00}$ modes at 775 nm (dashed lines, right axis). The green, red, blue and orange colors correspond to MPM and QPM with silicon oxide cladding and air cladding, respectively. With silicon oxide (air) cladding, the waveguide width is 1000 (1177) $\mu$m, respectively. Note that in \textbf{(b)}, the linear fit used to extract $\alpha $ is applied in the range 984 nm $\le w \le$ 1100 nm (1080 nm $\le w \le$ 1300 nm) for the MPM with silicon oxide (air) cladded waveguides and in \textbf{(c)} in the range 405 nm $\le e \le$ 440 nm for the MPM with air cladding.}
    \label{fig:simulation_tolerances}

\end{figure*}

Fig. \ref{fig:birds} d) shows the simulated SHG conversion efficiency as a function of the poling depth, $d$, for MPM between the second harmonic TE$_{01}$ mode and the fundamental TE$_{00}$ mode of the pump, and for QPM among fundamental TE$_{00}$ modes, for the cross-section show in Fig. \ref{fig:birds} c). The transverse component of the electric field at the different poling depths are shown in Fig. \ref{fig:birds} e). We define the SHG conversion efficiency ($\eta$) at phase matching and in the undepleted pump regime, i.e. assuming a constant pump power $P_{FH}$, as:
\begin{equation}
  \eta =   \frac{P_{SH} }{P_{FH}^2L^2 } =   \frac{2(\omega_{FH} \kappa_{SH})^2      }{ n^2_{FH} n_{SH}\epsilon_0 c^3}     \frac{  \iint e^*_{SH}(x,y)e_{SH}(x,y)\;dx\;dy  }{ (  \iint e^*_{FH}(x,y)e_{FH}(x,y)\;dx\;dy  )^2},
\end{equation}
where $P_{SH}$ is the power generated at the second harmonic, $L$ the length of the electrodes, $\omega_{FH}$ the frequency of the pump at the fundamental harmonic (FH), $\epsilon_0$ the vacuum permittivity, $c$ the speed of light in vacuum, $e_{FH}$ ($e_{SH}$) the electric field component of the fundamental (second) harmonic mode and $n_{FH}$ ($n_{SH}$) the effective indices of the fundamental (second) harmonic mode. The nonlinear coupling coefficient of SHG is defined as
\begin{equation}
    \kappa_{SH} =  \frac{\iint  \chi^{(2)} (x,y)e^*_{SH}(x,y)e_{FH}(x,y)e_{FH}(x,y)\;dx\;dy  }{  \iint e^*_{SH}(x,y)e_{SH}(x,y)\;dx\;dy }.
\label{Eq:kappa}
\end{equation}
It is expressed as an overlap between the mode profiles of the SH and FH in the $(x,y)$ plane. The integral at the numerator is only non-zero in the waveguide's core. For QPM we assume periodic poling with 50 \% duty cycle along the length of the waveguide, while we consider constant layer-poled poling in the MPM case, i.e. fully poled along the waveguide length as shown in Fig. \ref{fig:birds} b).  We define the nonlinear susceptibility as: 
\begin{equation}
   \chi^{(2)}(x,y) =  
\begin{cases}
      \beta d_{33} & \text{if $y>d$ }\\
      -\beta d_{33}  &  \text{if $y<d$}
    \end{cases}  
\end{equation}
where $y$ is the vertical coordinate of the waveguide's cross-section, $\beta = 1$ ($2/\pi$) for MPM (QPM) and $d_{33}= 25$ pmV$^{-1}$ is the highest coefficient of the tensor of the nonlinear susceptibility of lithium niobate \cite{shoji1997absolute}. Note that for QPM we consider $ \chi^{(2)}(x,y) = 0$ for $y>d$, to exclude the non contributing area, as illustated in Fig. \ref{fig:birds} e) with the blurred area. 

The maximum QPM efficiency is reached when the cross-section is uniformly poled, while the highest MPM efficiency happens when $d$ is at half the waveguide height ($h$). This is expected as, intuitively, the two lobes of the TE$_{01}$ mode occupy about half the waveguide height each, and they have the same magnitude. Also, the maximal MPM conversion efficiency is about 1.6 times higher than in QPM, though the mode overlap between the pump and the SH modes is higher in QPM. The reason lies in the lack of the $(2/\pi)^2$ factor, which generally lowers the conversion efficiency of QPM compared to MPM schemes.

\begin{figure*}[ht] 
\centering
  \begin{tabular}{cc}
    \includegraphics[width=0.45\linewidth]{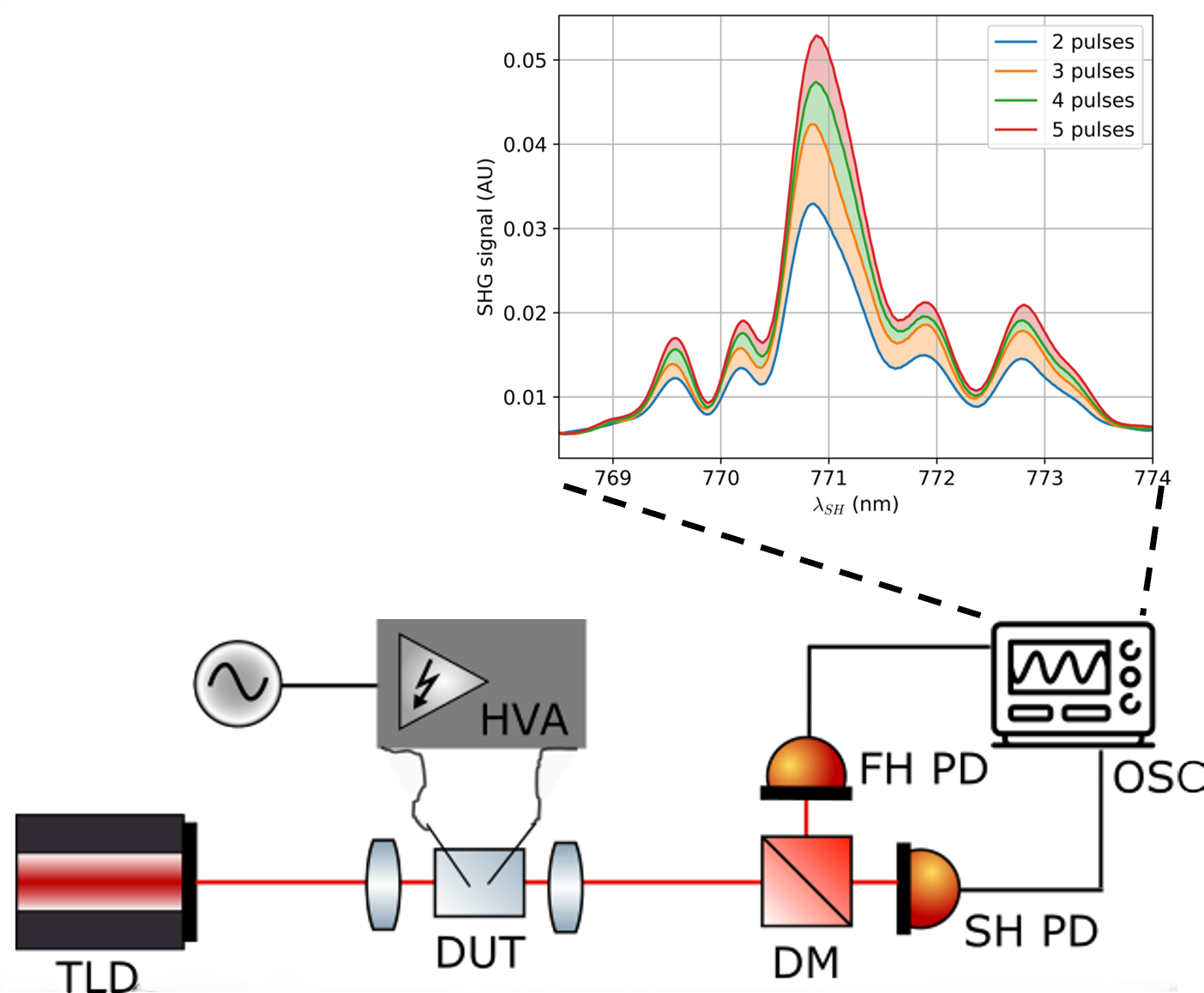} &
    \includegraphics[width=0.5\linewidth]{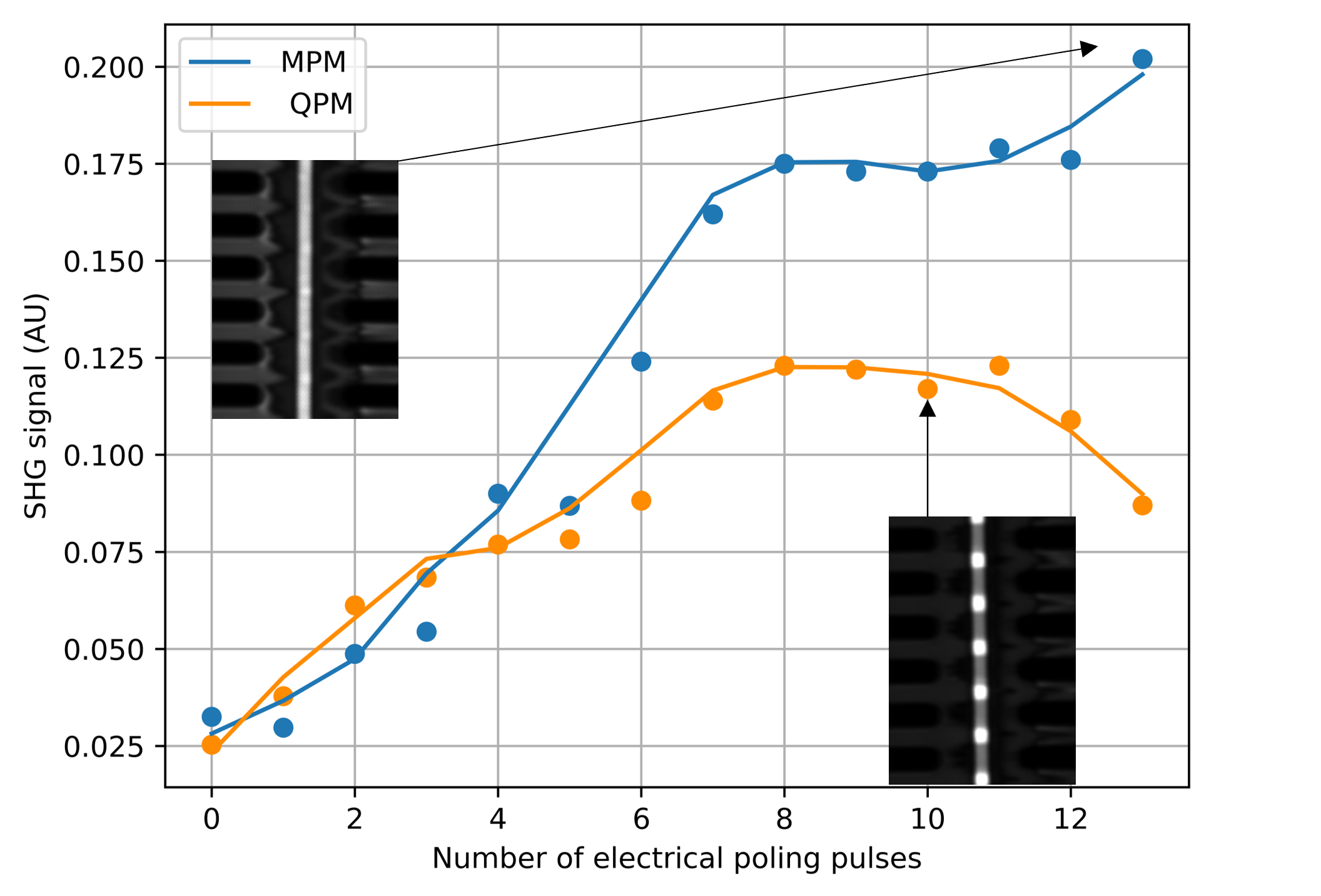}  \\
    \textbf{(a)} &  \textbf{(b)}  \\[6pt]
    \end{tabular}
   
      \begin{tabular}{cc}
    \includegraphics[width=0.4\linewidth]{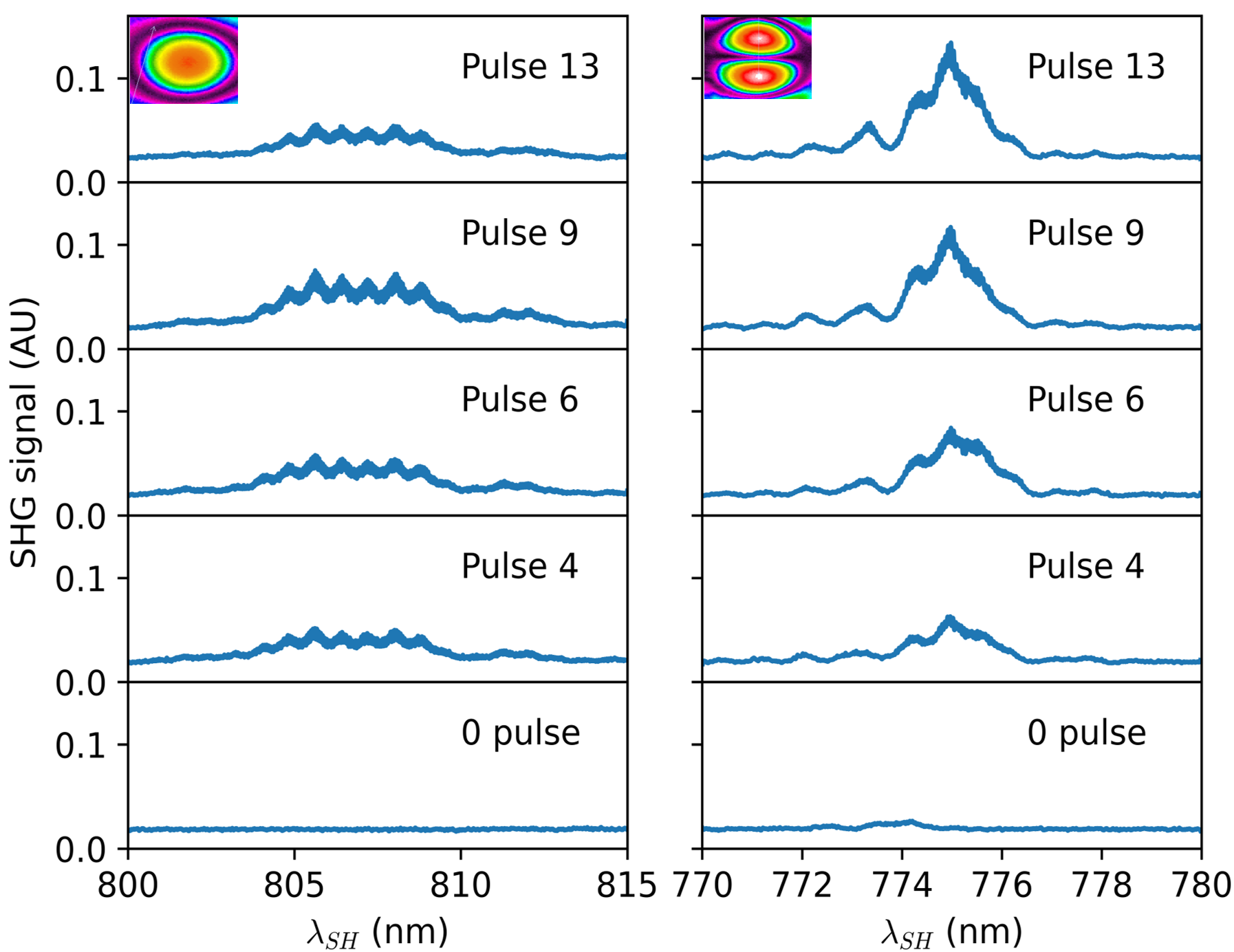} & 
     \includegraphics[width=0.45\linewidth,height=\textheight,keepaspectratio]{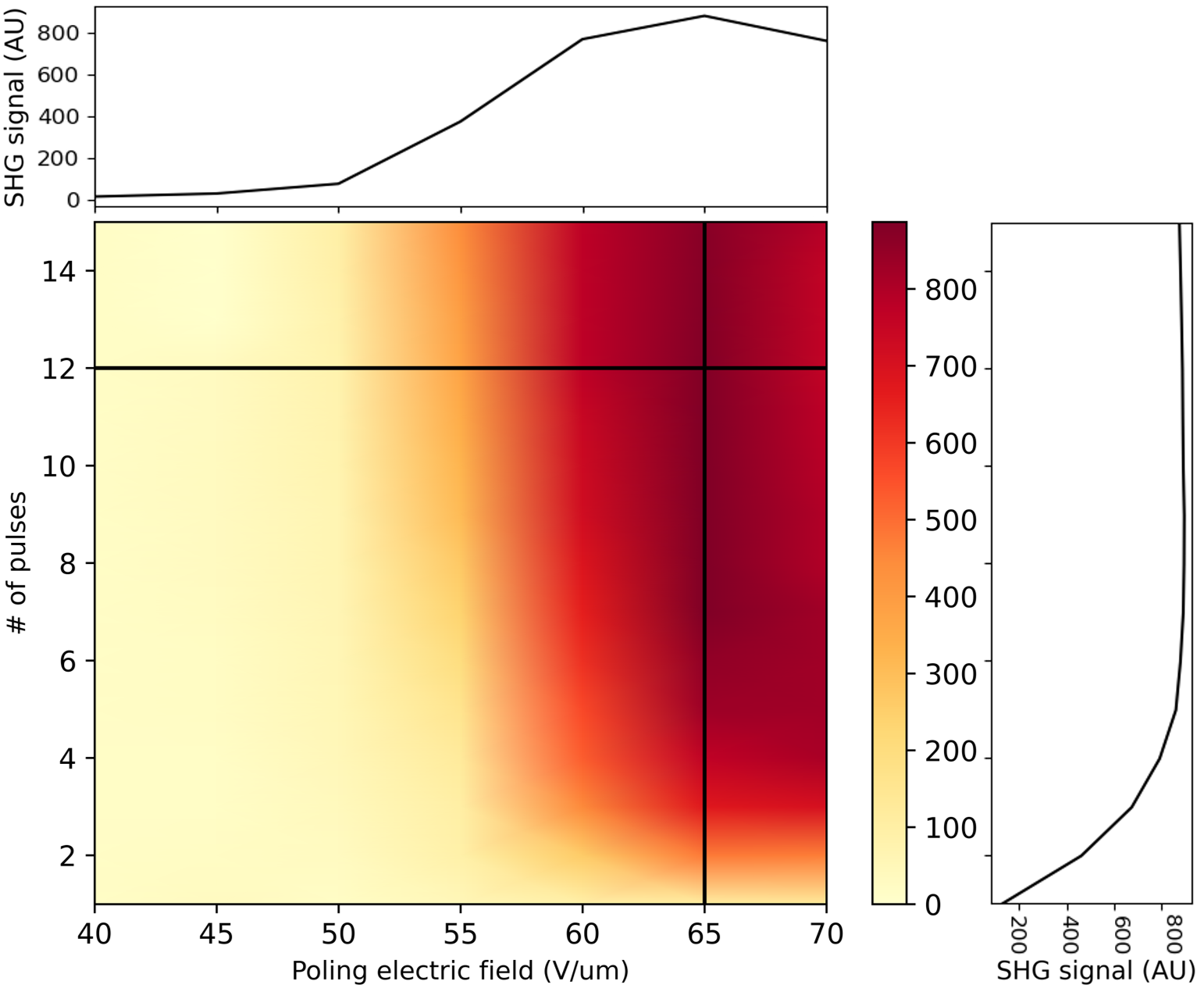}   \\
    \textbf{(c)} &  \textbf{(d)}  \\[6pt]
    \end{tabular}
    \caption
    { \textbf{(a)} Setup for electric-field poling optimization with real-time monitoring of the SHG. (TLD): tunable laser diode, (DUT): device under test, (DM): dichroic mirror, (FH/SH PD ): first /second harmonic photodetector, (OSC): oscilloscope, (HVA): high voltage amplifier. \textbf{(b)} Evolution of the SH power of the TE$_{00}$ mode and of the TE$_{01}$ mode as we apply pulses. Inset: two-photon microscope images taken after 8 and 13 pulses. \textbf{(c)} SH spectrum of the TE$_{00}$ mode (left) and TE$_{01}$ mode (right) at different number of pulses. Inset: detected profiles of the SH modes for QPM and MPM. \textbf{(d)} Optimization map of the MPM SH as a function of the applied electric field and number of pulses. The right (top) panels correspond to the SH profiles for a fixed electric field (number of pulses), indicated by the black lines.}
\label{fig:poling_live_monitoring}
\end{figure*} 

\subsection{\label{sec:tolerances_theory}Sensitivity of phase matching wavelength}
Fig. \ref{fig:simulation_tolerances} a)-d) show simulations of the phase matched pump wavelength, $\lambda_p$, in a SHG process as a function of the waveguide dimensions, both for QPM between fundamental modes and for MPM with SHG on the TE$_{01}$ mode. The green and red curves correspond to MPM and QPM in waveguides having a silicon oxide (SiO$_2$) cladding, respectively, while the blue and orange curves represent MPM and QPM in air cladded waveguides. In case of QPM, the poling period is fixed to have phase matching at $\lambda_p $ = 1550 nm for a cross-section featuring height ($h$) of 600 nm, etch depth ($e$) of 400 nm, side-wall angle ($\theta$) of 30$^\circ$, and width ($w$) of 1177 nm and 1000 nm for air- and SiO$_2$-cladded waveguides, respectively. We choose $w = 1177$ nm, such that $\lambda_p $ = 1550 nm for the MPM. For a variation of the parameter $x$, the sensitivity is defined as $\alpha = \frac{d\lambda_p}{dx}$.

We first note that $\lambda_p$ is almost one order of magnitude more sensitive to $h$ compared to $w$ and $e$, both for QPM and MPM. We also see that, in air (SiO$_2$)-cladded waveguides, the sensitivity of the QPM compared to MPM is 5.4, 9.4, 11.5 and 6.1 (3.5, 6.9, 3.4 and 4.0) times higher as a function of $h$, $w$, $e$ and $\theta$, respectively. This shows that MPM is more robust against fabrication uncertainties than QPM and that SiO$_2$-cladded waveguides are more tolerant. Fig. \ref{fig:simulation_tolerances} b) shows a linear range of $\lambda_p$ as a function of $w$ for the MPM with air (SiO$_2$)-cladding, which can be used for tunability, as well as a sweet spot at $w$= 1000 nm (940 nm), allowing low sensitivity to variations of $w$. 
Due to the shape of the TE$_{01}$ mode, we suspect that it might leak in the slab for certain dimensions, especially for shallow $e$. To catch this leaky behaviour, we add perfectly matched layers (PML) boundary conditions, that cross the slab along the simulation window. PML provide a lower bound of the propagation losses as they do not include scattering losses \cite{hu2009understanding,boes2019improved}. Fig. \ref{fig:simulation_tolerances} c) shows $\lambda_p$  for MPM and QPM (solid lines) and the propagation losses due to leakage in the slab (non-guidance) of the TE$_{01}$ and TE$_{00}$ modes at 775 nm in dashed lines for air (SiO$_2$)-cladded waveguides, as a function of $e$. We see that the propagation losses dramatically increase as $e$ gets shallower, reaching the dB/cm level for $e <$ 387 nm ($e <$ 400 nm) with air (SiO$_2$)-cladded waveguides.

\vfill

\begin{figure*}[ht]
\centering
    \begin{tabular}{cc}
    \includegraphics[width=0.45\linewidth]{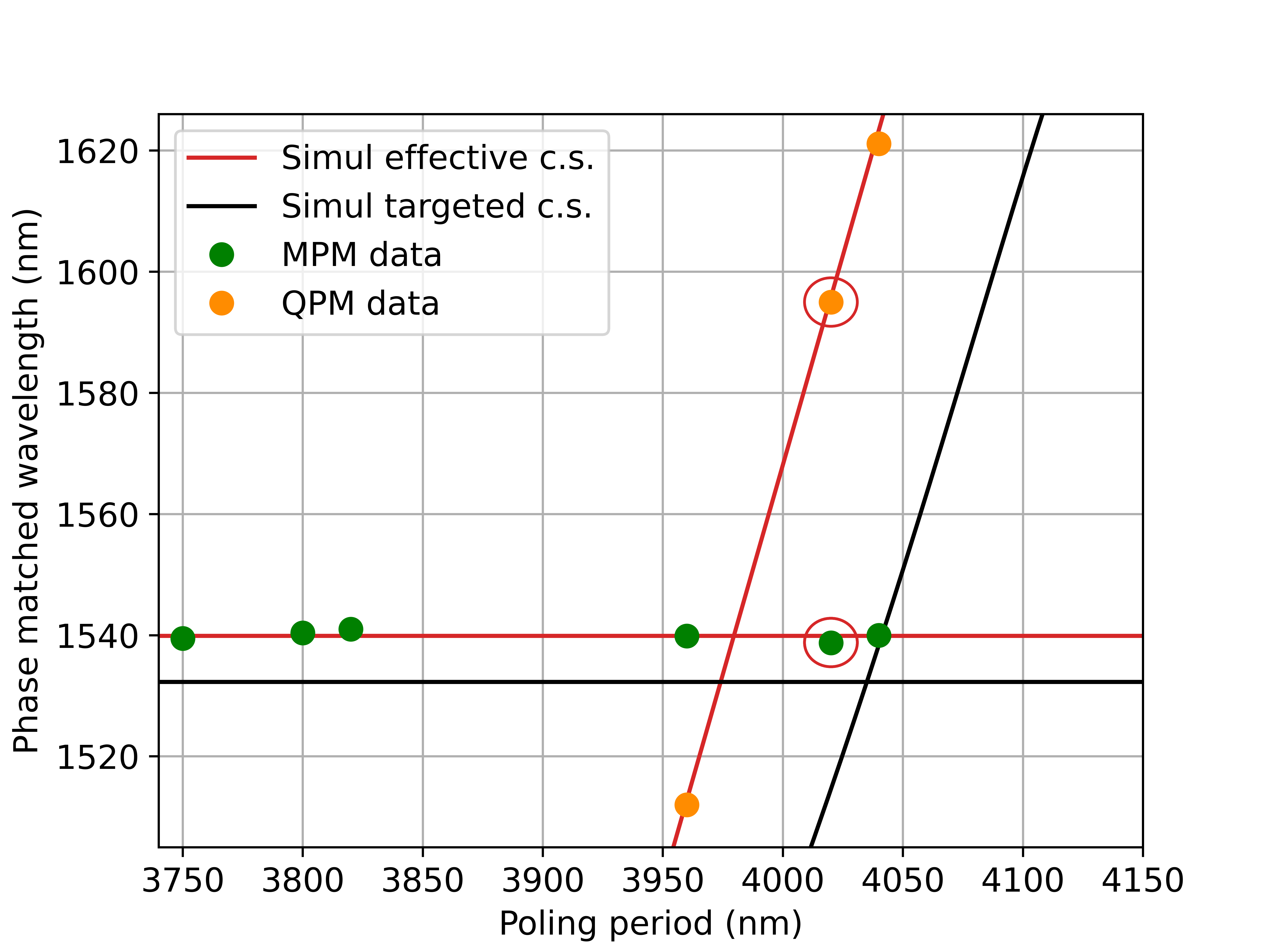} &
      \includegraphics[width=0.45\linewidth]{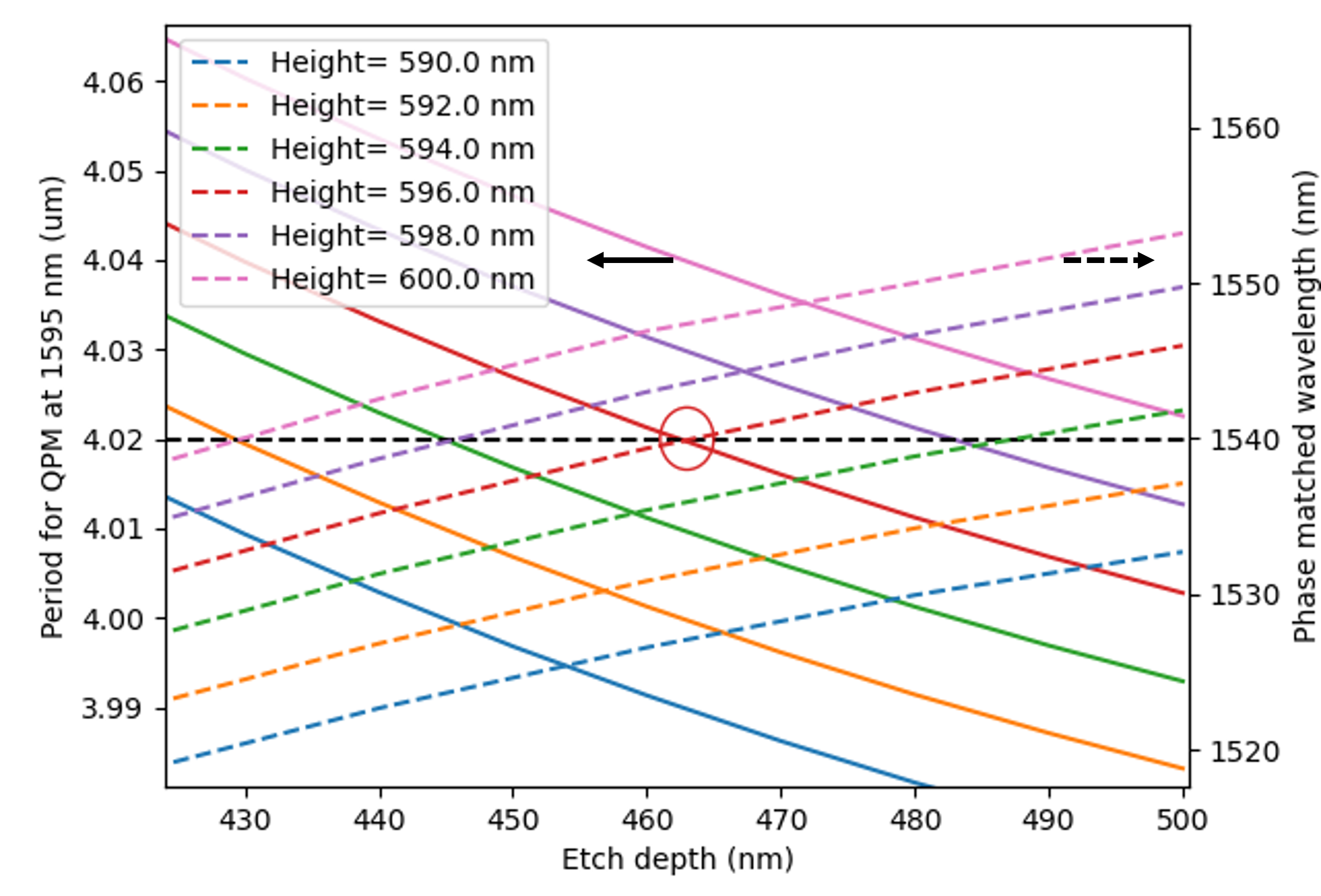} 
         \\
        \textbf{(a)} & \textbf{(b)} \\[6pt]
    \end{tabular}
    
    \begin{tabular}{cc}
 \includegraphics[width=0.45\linewidth]{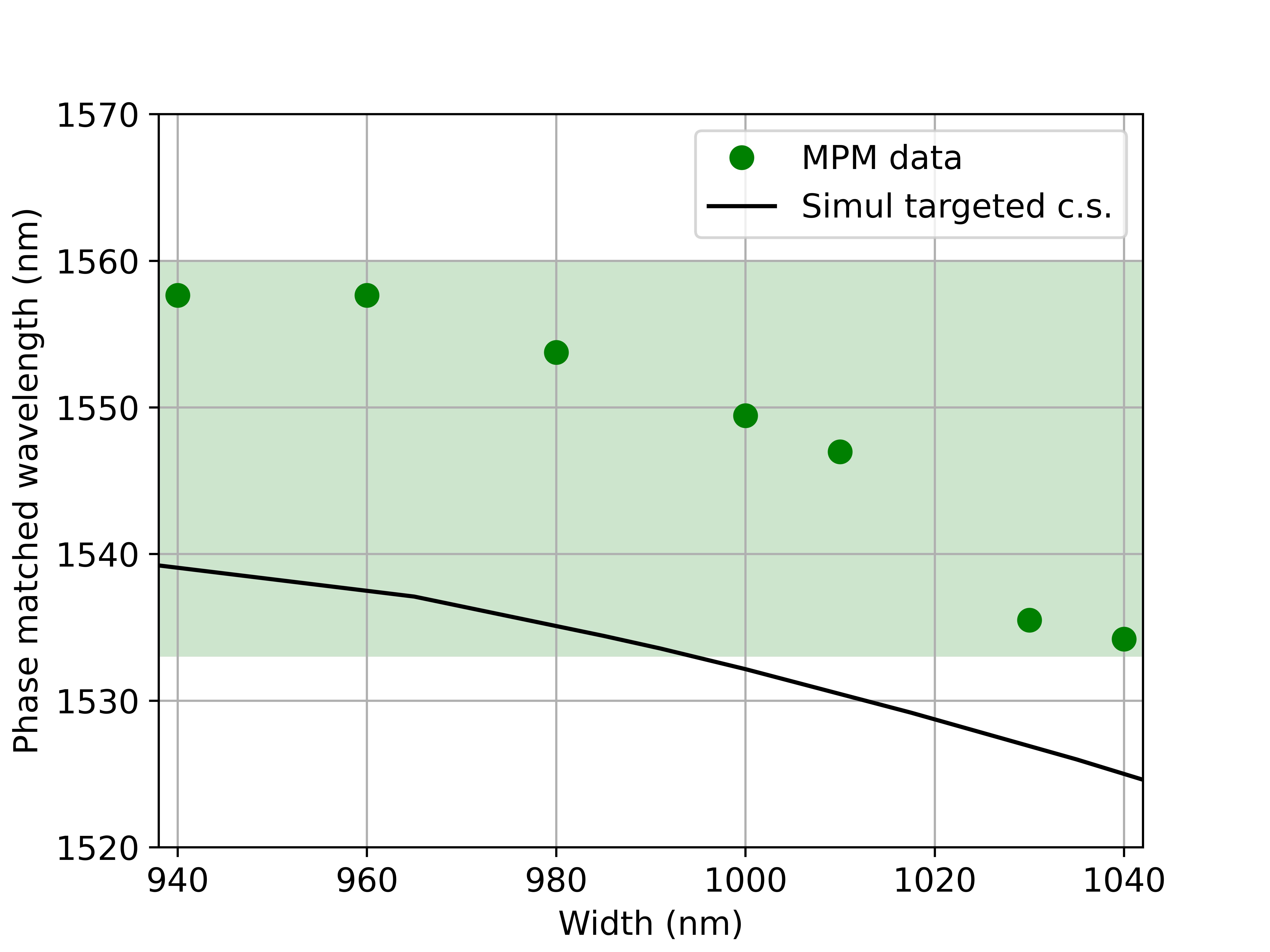} 
\\
        \textbf{(c)} \\[6pt]
    \end{tabular}
   
    \caption
    { \textbf{(a)} Detected phase matched wavelength of the FH  for QPM (orange dots) and MPM (green dots) schemes as a function of the poling period. The black curve corresponds to simulations with the targeted cross-section while the red curve corresponds to simulations with the waveguide dimensions matching the experimental data surrounded with red circles ($h$= 596 nm, $e$= 463 nm and $\theta = 30^\circ$). \textbf{(b)} QPM poling period for FH phase matched at 1595 nm (solid lines, left axis) and phase matched wavelength in MPM scheme (dotted lines, right axis) as a function of the etch depth and for different waveguide’s thickness. The left y axis scale is shifted to align the poling period required for QPM the FH at 1595 nm with the phase matched FH for MPM at 1540 nm, in agreement with the experimental observation of \textbf{(a)}. \textbf{(c)} Detected phase matched wavelength of the FH for MPM as a function of the waveguide width.
    }
\label{fig:tolerances_EXp}
\end{figure*}

\section{\label{sec:Experimental_results}Experimental results:\protect\\  }

\subsection{\label{section_samples}Waveguide fabrication}
The devices are fabricated in a wafer scale standard foundry process at CSEM \cite{csem_foundry}. The fabrication technology is based on commercially available TFLN on insulator wafers which consist of a stack with 600 nm thick mono-crystal X-cut LN layer on top of a 4.7 µm buried thermal oxide (BOX) layer. Waveguides are patterned by etching LN using ion milling, while layers are protected by a SiO$_2$ cladding. We pattern rounded tip gold electrode fingers with periods about 4 $\mu$m, 50 \% duty cycle and a gap of 8 $\mu$m after waveguide definition, to allow for the electric-field poling of the waveguides. Then, the waveguides are cladded with a SiO$_2$ layer and, optionally, this cladding can be selectively removed to access the electrodes’ pads and to obtain air-cladded waveguides.

\subsection{Poling optimization by real-time monitoring}
We perform electric-field poling at chip-level after fabrication, on etched waveguide. The setup is illustrated in Fig. \ref{fig:poling_live_monitoring} a), it allows to perform electric-field poling of the waveguides while monitoring the SH signal generated on the chip in real-time. Light from a tunable continuous wave laser (Santec TSL 510), delivered through a polarization-maintaining fiber, is collimated and passes through a high NA lens to be coupled into the chip. The output of the chip is collected and collimated through a second lens. A dichroic mirror is used to separate the SH from the FH which are sent on two separated photodetectors (PDA36A-EC and PDA10CS-EC), connected to an oscilloscope. We apply high voltage pulses to the electrodes using a waveform generator (Siglent SDG 1032 X) and a high voltage amplifier (TREK model 2220). We scan the wavelength of the laser while applying the poling pulses and we monitor the effect on the SH peak on the oscilloscope. This procedure allows to optimize the parameters and the number of pulses. 

We select a waveguide with a poling period of 4040 nm and a width of 1000 nm, allowing to obtain SHG both by MPM and QPM in the scanning range of our laser. We pole it while monitoring the SHG after each electric field pulse. MPM and QPM occur at different wavelengths, allowing us to separately monitor their evolution. We start with pulses with low electric field (55 V/um) to be able to see the evolution of the QPM due to the increase of the duty cycle. Fig. \ref{fig:poling_live_monitoring} b) shows the evolution of the SH signals as a function of the number of pulses revealing that the maximum SHG obtained by QPM is reached at about 10 pulses, while the MPM SH is still growing after 13 pulses. Note that the electric field was increased to 60 kV/um for the last point measurement, showing that the MPM is still further increasing. The insets show two-photon microscopy (TPM) images of the waveguide , confirming aperiodic poling  after 10 pulses and a linear poling after 13 pulses. The SHG spectra of points corresponding to 0, 4, 6, 9 and 13 pulses in Fig. \ref{fig:poling_live_monitoring} c), for both QPM and MPM, together with the beam profile of the SH mode detected at the waveguide output, which confirms the excitation of TE$_{00}$ and TE$_{01}$ modes for QPM and MPM, respectively.

It has to be noticed that, according to the simulations in Fig. \ref{fig:birds} a), the maximum SHG via QPM is likely limited by the height of the slab (about 200 nm). As the number of poling pulses increases, the waveguide gets overpoled, meaning that the poled domains merge along the waveguide's length. While, the SH in the TE$_{00}$ mode decreases, as the poled domains are deviating from the ideal 50 \% duty cycle for QPM , the SH in the TE$_{01}$ mode continues to grow, in agreement with the model for MPM which predicts the highest efficiency for a layer-poled slab gives.

Now, to find the optimal poling parameters for MPM, we repeatedly pole and de-pole another waveguide to obtain the SHG map shown in Fig. \ref{fig:poling_live_monitoring} d). We apply 15 square pulses of 1 ms duration at a given voltage, and we record the SHG after each pulse. Then, we erase the poling by applying a long pulse with reversed polarity until the SH peaks completely disappear on the oscilloscope. We then increase the pulse voltage and repeat the previous steps. As previously mentioned, the optimal MPM SHG is expected for a ratio $d/h =$ 50 \% and a poling duty cycle of 100 \%. With electric field of 55 V$\mu$m$^{-1}$ and below, we see that the SHG does not significantly grow even after all the 15 pulses. With higher electric field, the SH signal grows much faster with the number of pulses, reaching the maximum value after 12 pulses at 65 V$\mu$m$^{-1}$ (solid black lines). 

\subsection{\label{sec:tuing_phase_matching}Tuning of the phase matching wavelength}

As we have numerically shown in Section \ref{sec:tolerances_theory}, $\lambda_p$  exhibits a large sensitivity to the different dimensions of the waveguide's cross-section. Among these parameters, $w$ is the most controllable one, with resolution around $\pm 10$ nm and it can be set by design. Also, the poling period can be intentionally tuned to change $\lambda_p$ in QPM. Therefore, we tested several waveguides placed on two different chips. Each chip covers an area of 5x5 mm$^2$. On the first (second) chip, we vary the poling period (width). Also, we considered just SiO$_2$ cladded waveguides, as the addition of silicon oil, required to avoid air breakdown during poling, shifts $\lambda_p$ out of the tuning range of our laser in air-cladded waveguides, and the SH cannot therefore be detected in real time. Fig. \ref{fig:tolerances_EXp} a) shows the recorded $\lambda_p$  for both MPM and QPM of several nominally identical waveguides with $w$ = 1000 nm, with respect to the poling period. In case of QPM, the large slope indicates high sensitivity of $\lambda_p$ while, as expected, the MPM SH does not depend at all on the poling period. Overall, the data well-follow the predicted sensitivity but show an offset compared to the target cross-section, indicated by the black curves. To better understand this offset, we performed additional simulations by varying the waveguide’s cross section to find the set of parameters that can reproduce the experimental points. As many combinations of $h$, $e$, and $\theta$ can give the same $\lambda_p$ for a given phase matching scheme, we restricted the range of $h$ values to the ellipsometry measurement provided by the wafer’s supplier, which corresponds to $h$ = 595 $\pm$ 5 nm at this specific chip location, and we kept $\theta$ = 30$^\circ$, as designed. Moreover, we exploit the simultaneous detection of SH from MPM and QPM, to find the combination of $h$ and $e$ that could simultaneously provide QPM and MPM at the observed wavelengths and poling period. Fig. \ref{fig:tolerances_EXp} b) shows the periods corresponding to QPM at 1595 nm (solid lines, left axis) and the MPM $\lambda_p$ (dashed lines, right axis) as a function of the etch depth for different waveguide heights. The black horizontal line indicates a period of 4.02 $\mu$m and a MPM at 1540 nm, which correspond to two experimental points surrounded with red circles in Fig. \ref{fig:tolerances_EXp} a). As shown by the red lines in Fig. \ref{fig:tolerances_EXp} a), we could well-reproduce the experiments by setting $h$ = 596 nm and $e$ = 463 nm. Also, this plot clearly confirms the larger robustness of MPM compared to QPM. Indeed, the actual cross-section results in 80 nm off-set for $\lambda_p$ in case of QPM, while for MPM this value reduces to 6.6 nm. For this effective cross-section, the model predicts propagation losses of 3.3 $10^{-11}$ dBcm$^{-1}$ at 770 nm for the TE$_{01}$ mode due to field decay in the slab, which are negligible. In MPM, $\lambda_p$ cannot be independently tuned to the cross-section of the waveguide, while this can be achieved in QPM. Nevertheless, in Fig. \ref{fig:tolerances_EXp} c), we show that in practice we could well address selected $\lambda_p$ over about 30 nm inside the C-band by tuning the width of the waveguide by about 100 nm. This value is small enough not to noticeably change the mode guidance and the efficiency of the process.
In Fig. \ref{fig:tolerances_EXp} c), the experimental points well-reproduce the trend of the simulations (black curve), confirming the suitability of the lithographic precision on $w$ for this purpose.

 \begin{figure*}[ht]\centering
  \begin{tabular}{ccc}
     \includegraphics[width=0.337\linewidth,height=\textheight,keepaspectratio]{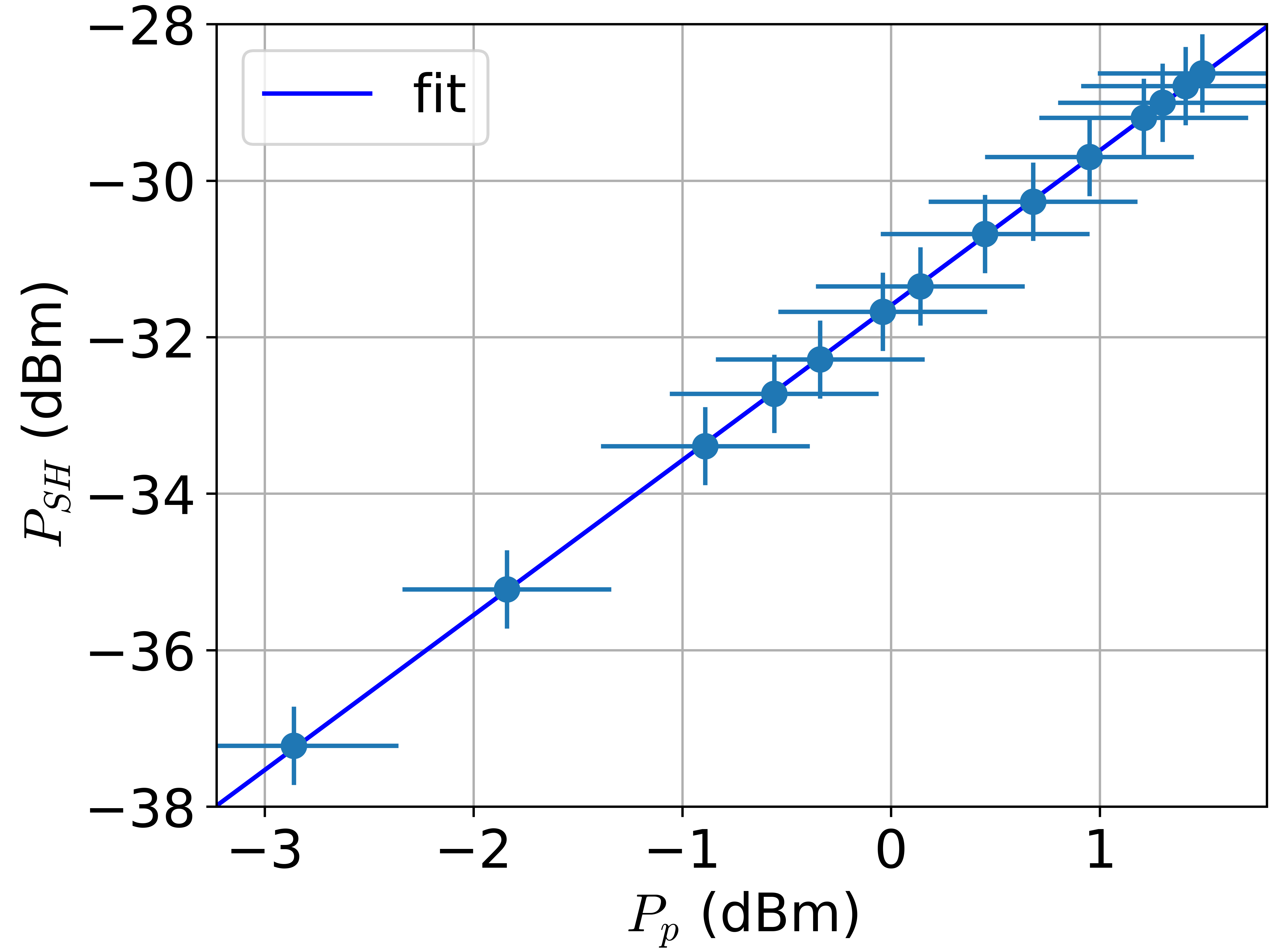} &
         \includegraphics[width=0.341\linewidth,height=\textheight, keepaspectratio]{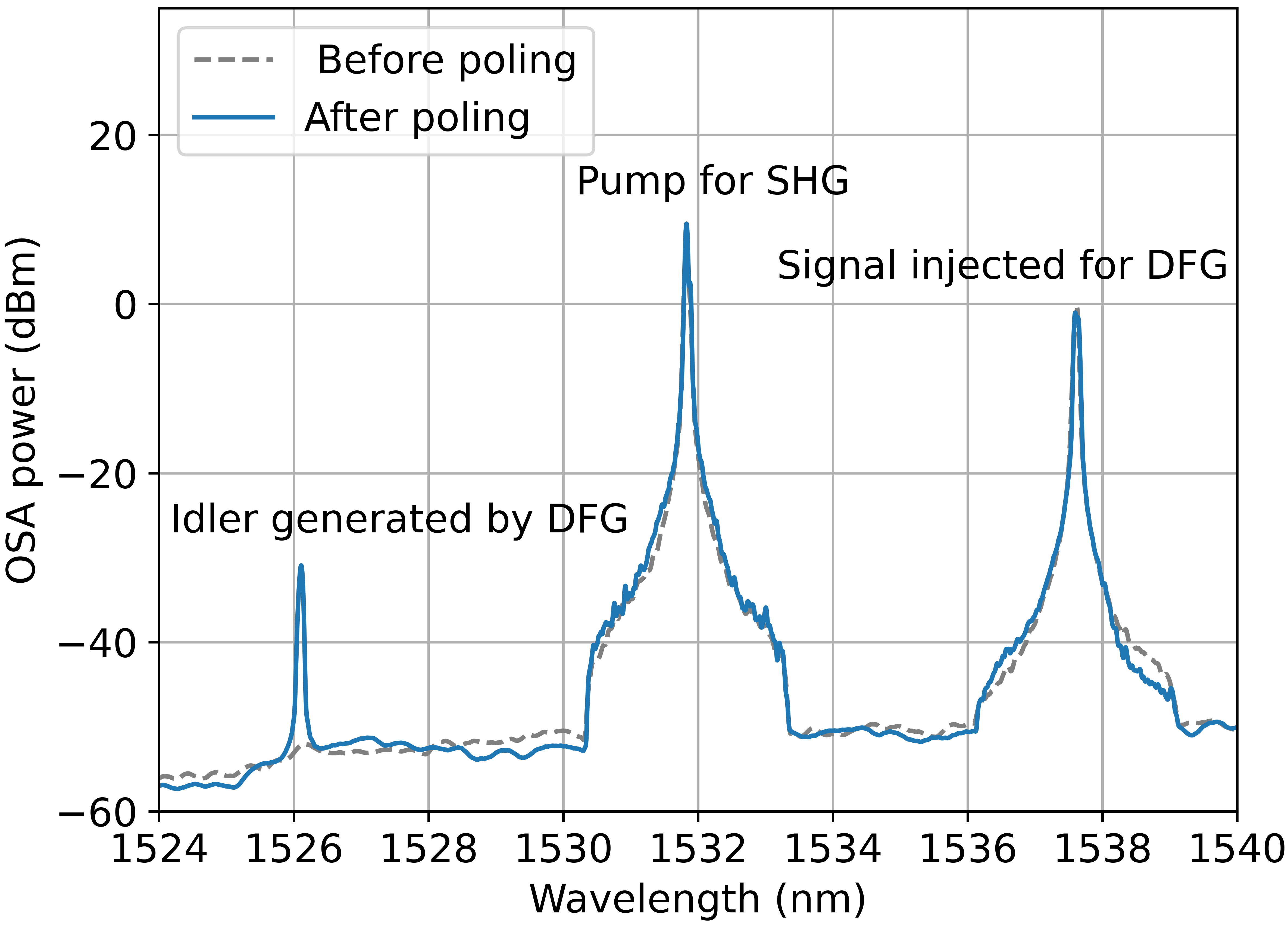}    &
      \includegraphics[width=0.329\linewidth,height=\textheight,keepaspectratio]{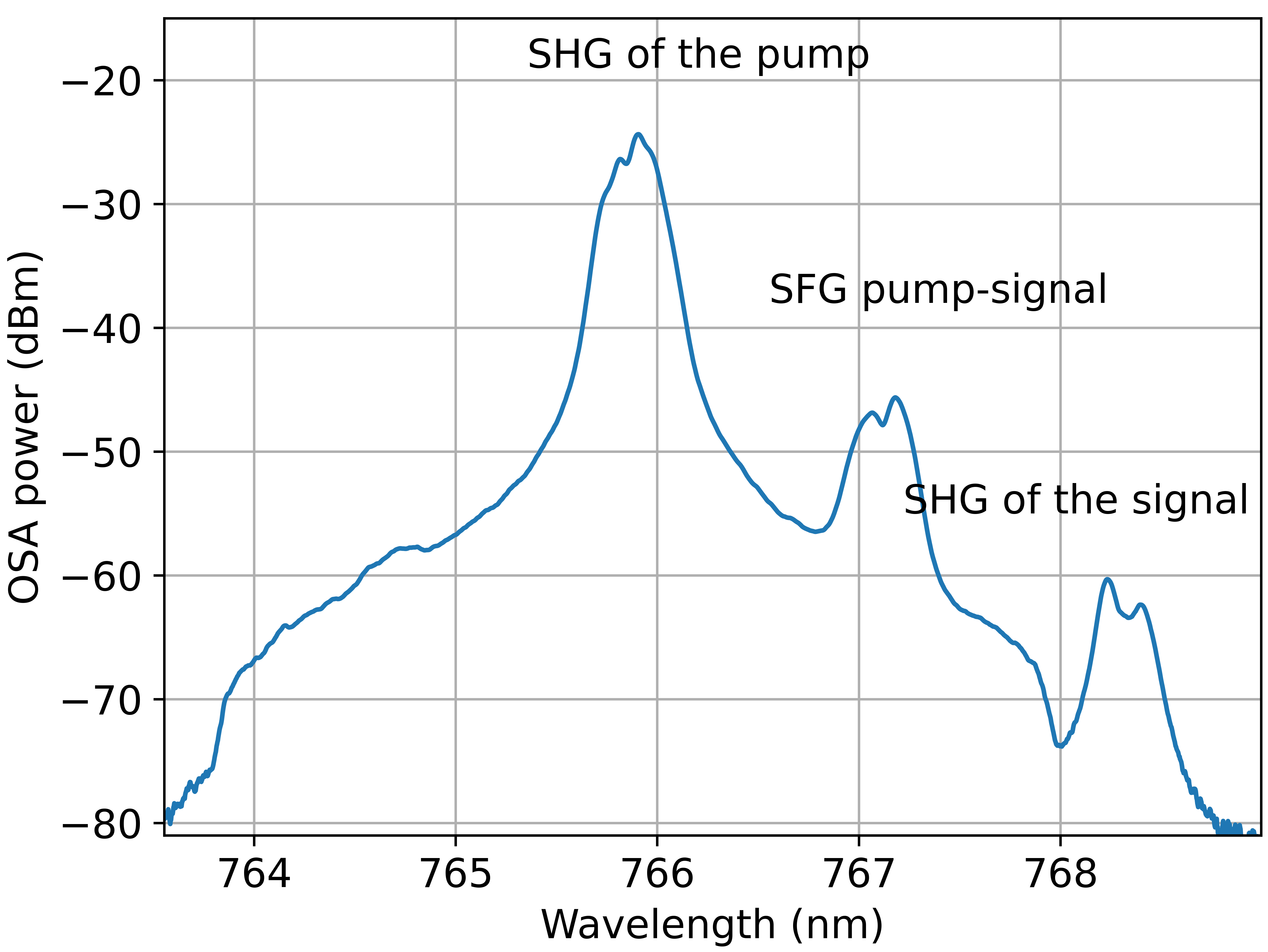} 
        
     \\
          \textbf{(a)} & \textbf{(b)}   & \textbf{(c)}  \\[6pt]
     \end{tabular} 
     \begin{tabular}{ccc}
    \includegraphics[width=0.4\linewidth,height=\textheight, keepaspectratio]{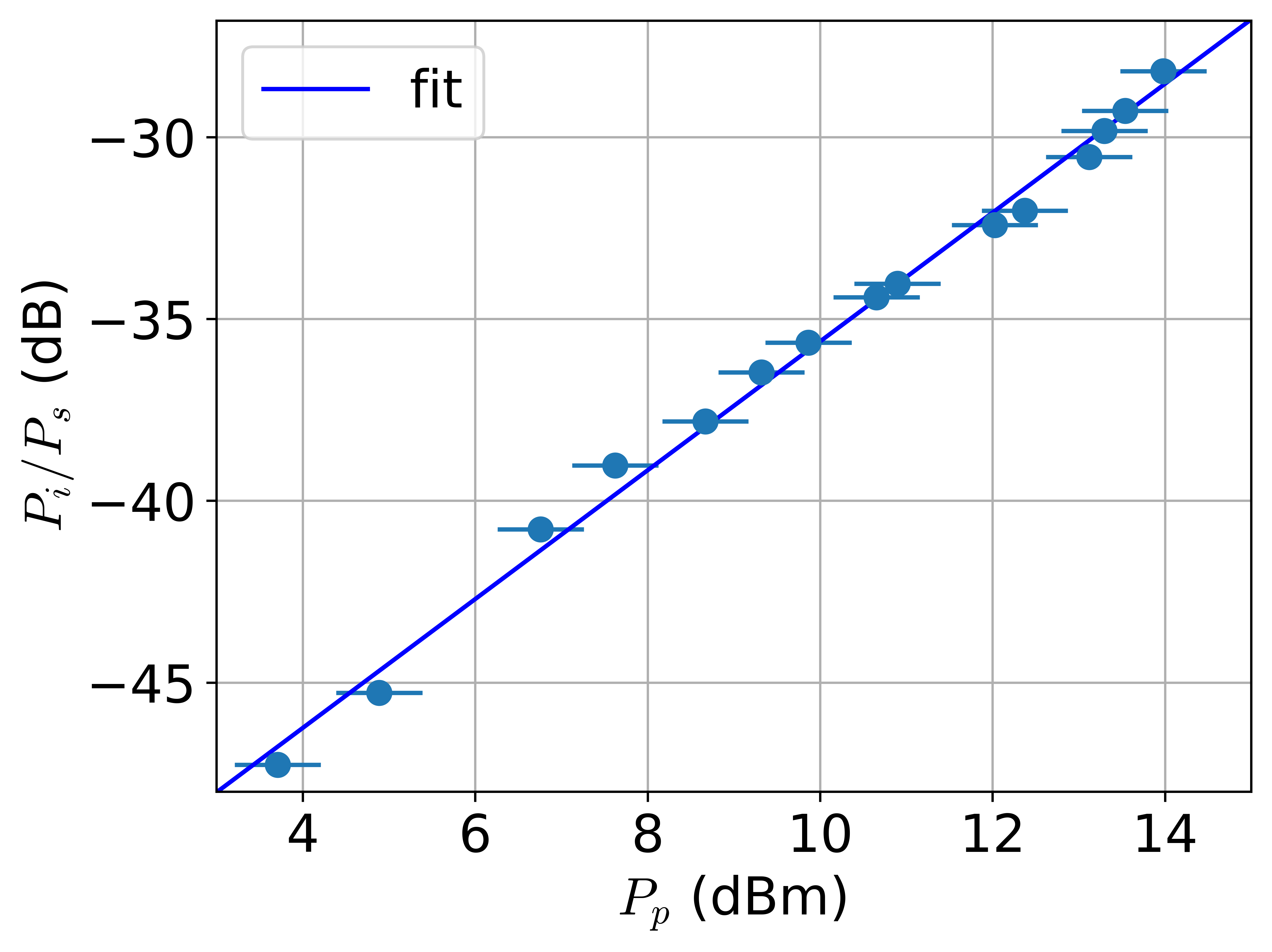}     &
        \includegraphics[width=0.41\linewidth,height=\textheight, keepaspectratio]{ 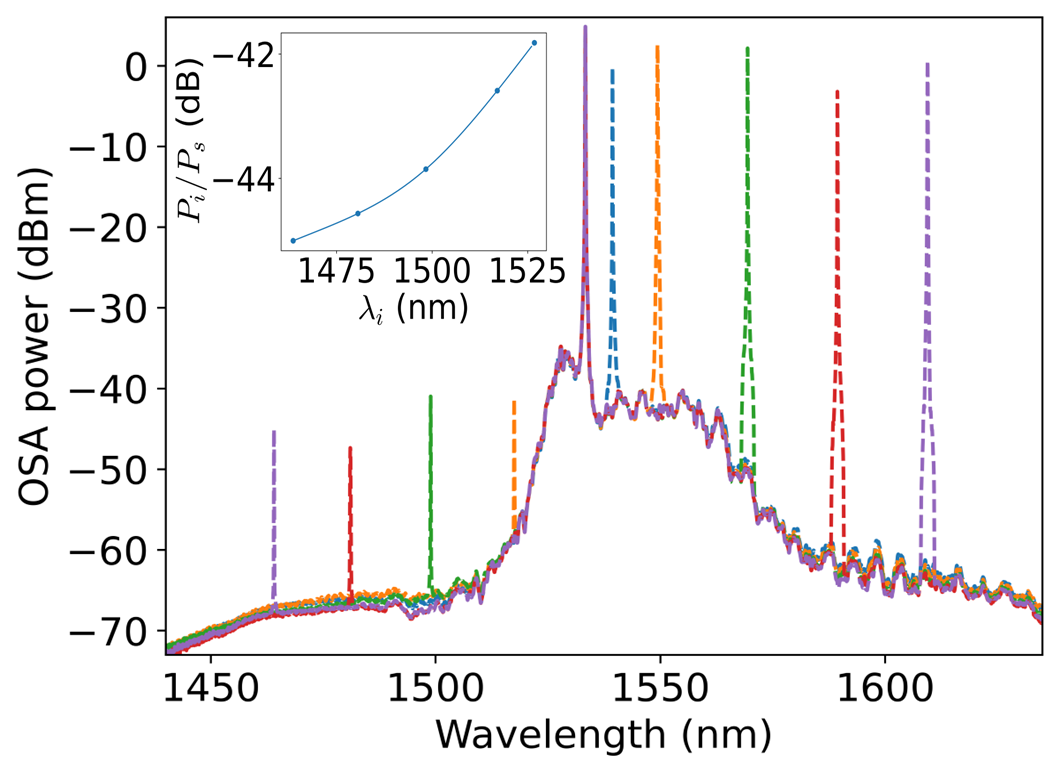}     
     
     \\
          \textbf{(d)} & \textbf{(e)}  \\[6pt]
     \end{tabular}
     \caption
     { \textbf{(a)} Estimated on-chip SH signal from MPM as a function of the estimated on-chip pump power. Recorded spectra on the OSA in \textbf{(b)} FH and \textbf{(c)} SH wavelength regions before (grey dotted lines) and after (blue lines) the poling of the waveguide. \textbf{(d)} Signal-to-idler ratio as a function of the on-chip pump power. \textbf{(e)} DFG spectrum for various signal wavelengths. Inset: Signal-to-idler ratio as a function of idler wavelength. }
 \label{fig:cascaded_SHG_DFG}
 \end{figure*}

\subsection{Interband and intraband frequency conversion processes with MPM}
We first experimentally assess the performance of MPM for SHG. The SiO$_2$-cladded waveguide used in the following experiment has a double inverse taper edge coupler \cite{he2019low} optimized for 1550 nm at the input, and a straight waveguide at the output. The taper-less output is intended to limit the outcoupling losses of the SH mode on the TE$_{01}$ mode. To estimate the conversion efficiency, we measure the output power of both the pump and of the SH by placing a photodetector (PD) immediately after the output facet of the chip. We assume no collection losses at the output for both the pump and the SH. Therefore, we attribute the measured transmission losses to the input coupling. We use a silicon PD for the SH, sensitive in the range of 200-1100 nm and a germanium PD for the pump. Considering that the electrodes of the waveguide are 4.4 mm long, we estimate a normalized SHG conversion efficiency of 430 \% W$^{-1}$cm$^{-2}$ from a single point measurement.  
Then, we place the two PD after a collecting lens and a dichroic   and we estimate the additional output coupling losses coming from the output lens and the dichroic to be $-1.5$ dB $\leqslant \alpha_p \leqslant -0.5$ dB for the FH and $-3.5$ dB $\leqslant \alpha_{SH} \leqslant -2.5$ dB for the SH. The estimated on-chip SH power as a function of the on-chip pump power is shown in Fig. \ref{fig:cascaded_SHG_DFG} a), where the uncertainty bars come from the estimation of the coupling losses. We see a slope of 1.98 in log-log scale, confirming the quadratic dependence of SH on the pump power. Moreover, we did not observe any sign of SHG saturation, meaning that we are still in the regime of undepleted pump. From the fit of the experimental points, we estimate a normalized SHG conversion efficiency of (360$\pm$ 90) \% W$^{-1}$cm$^{-2}$, in line with the single point measurement. 
This value is about one order of magnitude lower than the simulations and we mainly attribute this mismatch to the actual thinner slab and to non-uniform poling along the waveguide length. A further detailed discussion can be found in the supplementary material.

Then, we employed the SH generated through MPM as the pump for a second down conversion process within the same waveguide. The overall result is therefore an intraband frequency conversion module, working at the telecom band. For this purpose, we use an air-cladded waveguide featuring SHG via MPM occurring at 1531.8 nm, and with double inverse tapers at both input and output to optimize the power extraction in the telecom band. We inject a pump at the phase matched wavelength together with a lower power signal at 1537.5 nm. Two cascaded processes take place: SHG of the pump and DFG between the signal and the SH of the pump, generating an idler at $\lambda_i = (\frac{1}{\lambda_{SH}} -\frac{1}{\lambda_s} )^{-1}$, where $\lambda_s$ and $\lambda_{SH}= \frac{\lambda_p}{2}$ are the wavelengths of the signal and of the SH of the pump. We use the same setup as in Fig. \ref{fig:poling_live_monitoring} a), but we add a 3 dB fibered beam combiner at the input, and at the output we focus the light into a multi-mode fiber from the collimator. The output spectrum, recorded in an optical spectrum analyzer (OSA), is shown in Fig. \ref{fig:cascaded_SHG_DFG} b). We clearly see the generation of an idler, blue-shifted with respect to the pump wavelength by the same amount of the shift between the pump and signal. 
It has to be noted that, like the cascaded SHG and DFG processes, a phase matched four-wave-mixing (FWM) process, enabled by the third order nonlinearity of lithium niobate, can also generate an idler at $\lambda_i$. Therefore, to exclude any contribution from FWM, we tested the waveguide before electric-field poling, using the same setup. The output spectrum does not show any idler generation, for the power level employed, as shown by the dashed trace in Fig. \ref{fig:cascaded_SHG_DFG} b), confirming that the recorded idler is generated by the cascaded SHG-DFG process. After poling, three peaks appear on the spectrum in Fig. \ref{fig:cascaded_SHG_DFG} c). They correspond to the SH of the pump, to sum frequency generation between the pump and the signal and to the SH of the signal. 
We then record the ratio between the idler and signal power as a function of the estimated on-chip pump power. The data, shown in Fig. \ref{fig:cascaded_SHG_DFG} d) are obtained by reporting the power detected on the OSA, after subtraction of the background noise, and considering coupling losses in the multimode fiber of -0.9 dB and in the OSA of -4.36 dB. Given the negligible chromatic aberration of the employed optical elements at the chip output, we assume equal losses at the three wavelengths. 
We define a conversion efficiency of the overall cascaded processes as $\eta_{SHG-DFG} = P_i/P_sP^2_p$, where $P_i$, $P_s$ and $P_p$ are the on-chip powers of the idler, of the signal and of the pump in the telecom band). Fitting the data in Fig. \ref{fig:cascaded_SHG_DFG} d), we obtain $\eta_{SHG-DFG}=(280\pm 60)$ \%W$^{-2}$.
We then assess the bandwidth of the cascaded process. Contrarily to the SHG generation only, which features a bandwidth limited to the nanometer range, the DFG process has a much higher bandwidth \cite{koyaz2024ultrabroadband}. Fig. \ref{fig:cascaded_SHG_DFG} e) shows the idler efficiently generated for different signal wavelengths over a band >100 nm. 

\section{Conclusion and discussion}

We have shown that SHG can be generated by MPM in the TE$_{01}$ waveguide mode by selectively poling the lower part of TFLN waveguides, fabricated at wafer scale through a foundry process. We optimized the poling by monitoring the SHG signal during the application of the electric field pulses. The MPM approach does not depend on poling duty cycle and period, making the poling process much easier and enabling more reproducible conversion efficiency than standard QPM. We simulated the shift of the phase matched wavelength with respect to deviation to the nominal cross-section parameters, and we obtained that MPM is from 5 to 10 times less sensitive than QPM depending on the parameter considered, while exhibiting a theoretical conversion efficiency 1.6 times higher. We experimentally validated our simulations by varying the parameters that can be controlled by design, namely the waveguide width, the cladding vs option, and the poling period, for the QPM case. Finally, we experimentally demonstrated the use of MPM for interband frequency conversion, by SHG, and for intraband frequency conversion, by using the same waveguide to host two cascaded nonlinear processes: up-conversion by SHG and down- conversion by DFG. We estimated a SHG conversion efficiency of (360$\pm$ 90) \% W$^{-1}$cm$^{-2}$ and a conversion efficiency of the cascaded processes of $\eta_{SHG-DFG}=(280\pm 60)$ \% W$^{-2}$ over about 120 nm of bandwidth. This value is about $1000$ ($200$) times higher than telecom band intraband frequency conversion based on FWM in silicon nitride waveguides \cite{grassani2018second} (silicon photonics waveguides of similar length \cite{foster2006broad,borghi_nonlinear_2017}) and makes this approach a viable solution for all-optical frequency conversion in telecommunication networks. Also, the theoretical MPM SHG conversion efficiency would lead to conversion efficiency for the cascaded process of the order of $\sim10^4$ \%W$^{-2}$, comparable to what can be obtained in highly nonlinear ring resonators \cite{azzini_classical_2012}, although featuring a much higher bandwidth. Overall, such results indicate that layer-poled electric-field poling on TFLN waveguides performed at the back end of the process is a powerful and versatile method to obtain efficient and reproducible photonic integrated frequency converters fabricated at wafer scale.

\section*{Supplementary Material}
The supplementary material can be found in a separate pdf file.

\section*{ Acknowledgements}
This Bridge project is funded by the Swiss National Science Foundation (SNSF) (194693). 
\section*{Author declarations}
\subsection{Conflict of Interest}
The authors have no conflicts to disclose.

\section*{Data Availability Statement}
The data that support the findings of this study are available from the corresponding author upon reasonable request.

\section*{References}
\nocite{*}
\bibliography{MPM_Draft_v2_biblio}
\end{document}


\preprint{AIP/123-QED}

\title{Fabrication-tolerant frequency conversion in thin film lithium niobate waveguide with layer-poled modal phase matching\\
Supplementary material}

\author{O. Hefti}
 \altaffiliation[Also at ]{EPFL, Photonic Systems Laboratory (PHOSL), 1015 Lausanne (CH).}
\author{J.-E. Tremblay}%
\author{A. Volpini}
 \email{olivia.hefti@csem.ch}
\affiliation{ 
CSEM SA, Rue de l'Observatoire 58, 2000 Neuchâtel (CH).
}%

\author{Y. Koyaz}
\affiliation{%
EPFL, Photonic Systems Laboratory (PHOSL), Rte Cantonale, 1015 Lausanne (CH).
}

\author{I. Prieto}
\author{O. Dubochet}
\author{M. Despont}
\author{H. Zarebidaki}
\author{C. Caër}
\author{J. Berney}
\author{S. Lecomte}
\author{H. Sattari}%
\affiliation{ 
CSEM SA, Rue de l'Observatoire 58, 2000 Neuchâtel (CH).
}%

\author{C.-S. Brès}
\affiliation{%
EPFL, Photonic Systems Laboratory (PHOSL), Rte Cantonale, 1015 Lausanne (CH).
}%

\author{D. Grassani}
\affiliation{ 
CSEM SA, Rue de l'Observatoire 58, 2000 Neuchâtel (CH).
}%

\date{\today}
\maketitle



\subsection{\label{Efficiency discrepancy}Efficiency discrepancy discussion}
The measured conversion efficiency is about one order of magnitude lower than the simulation with a poling depth of 200 nm, the targeted slab height, up to which we expect the poling to be effective. Here we discuss the reasons that could explain this discrepancy.


\begin{figure*}[ht]

    \begin{tabular}{ccc}
    \includegraphics[width=0.35\linewidth]{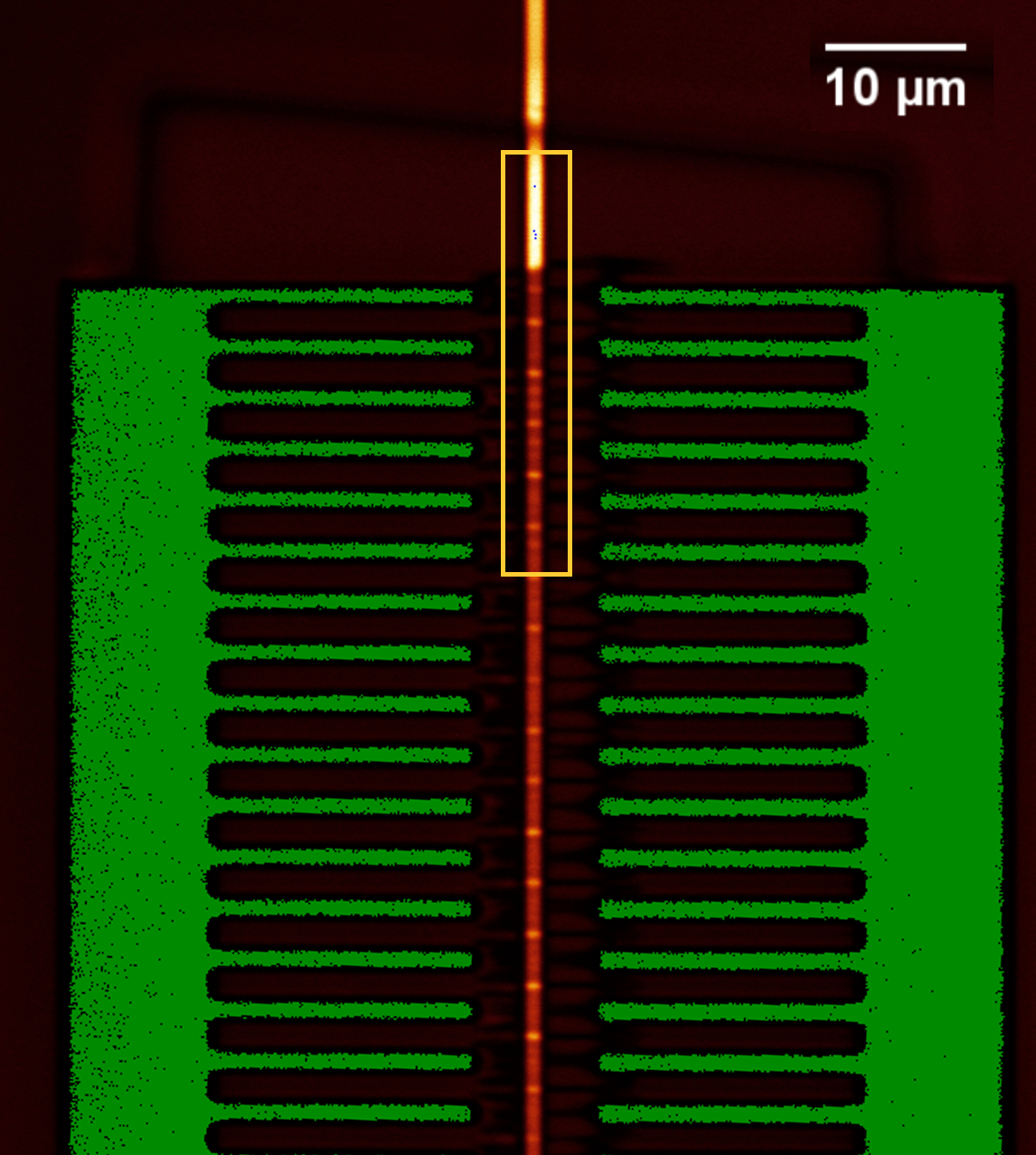} &
    \includegraphics[width=0.6\linewidth]{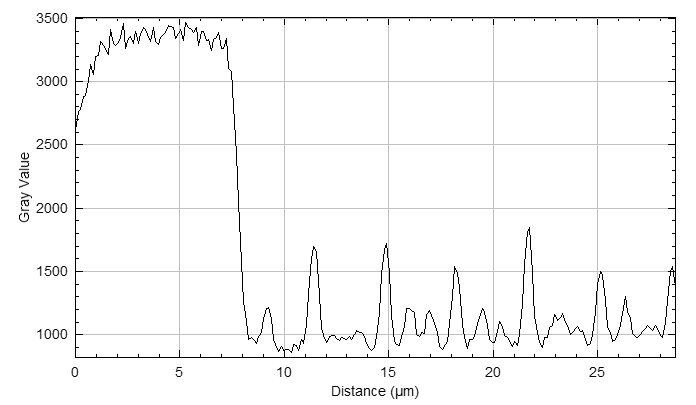} \\
    \textbf{(a)} &  \textbf{(b)}    \\[6pt]
    \end{tabular}
    \caption{ \textbf{(a)} TPM image showing the electrodes in green and the waveguide in orange. \textbf{(b)} Intensity along a longitudinal cut at the center of the waveguide in the region identified by the yellow box is shown in \textbf{(a)}. }
    \label{fig:TPM}
\end{figure*}

\begin{figure*}[ht]
    \includegraphics[width=0.7\linewidth]{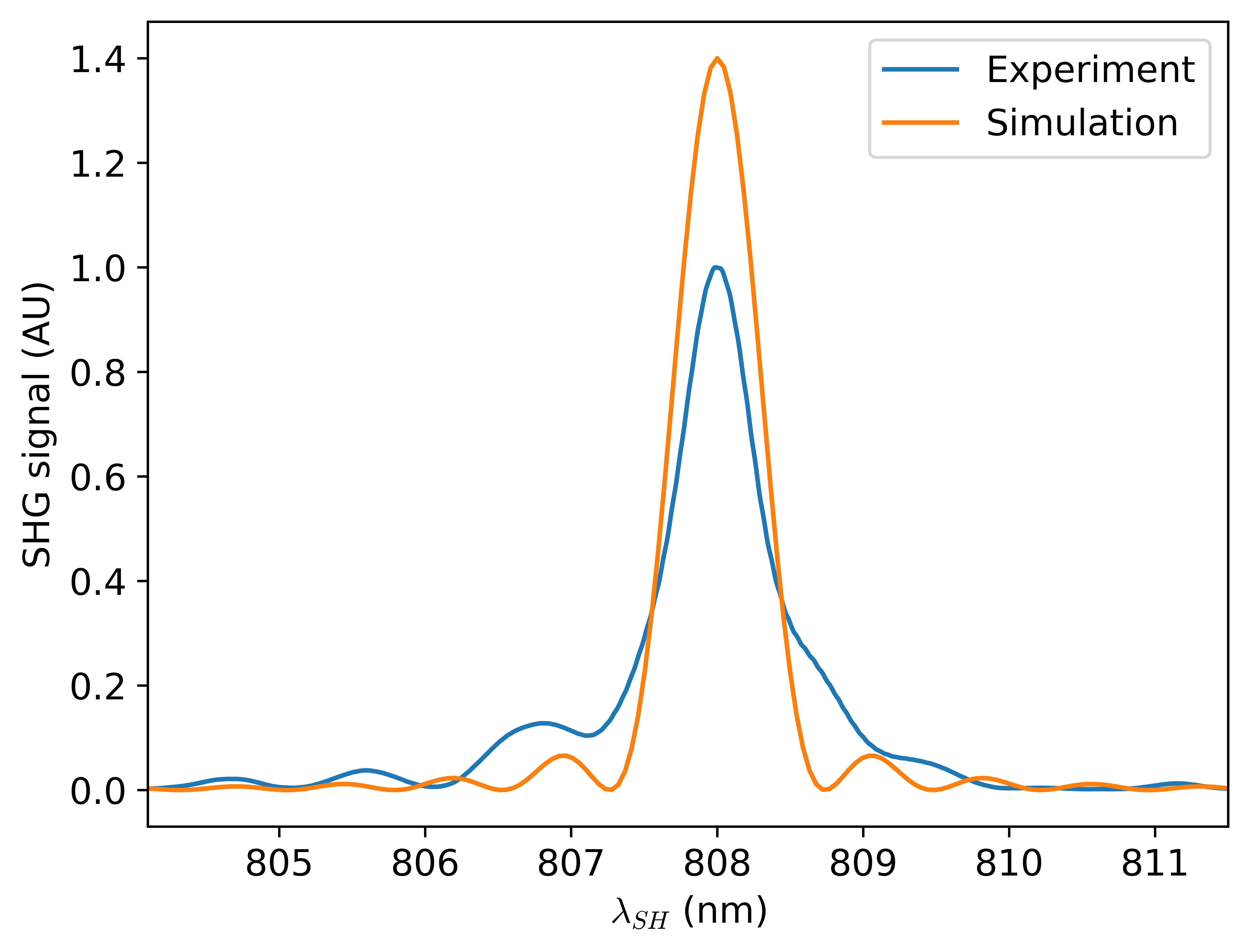} 
    \caption{Normalized experimental SHG spectrum (blue line) and simulated one (red line) rescaled to obtain the same integral area. }
    \label{fig:SHG_spectrum_gradient_dimensions}
\end{figure*}

First, we have seen in the subsection III C that simulation with an etch depth of 463 nm reproduces the phase matched wavelength of the waveguide considered. We can assume that this etch depth limits the poling depth to 20-30 \% of the initial waveguide thickness, which corresponds to an efficiency between 1000 and 3000 \% W$^{-1}$cm$^{-2}$ according to the simulation in Fig. 1 \textbf{(c)} in the main paper. TPM images also give insights on the poling depth, as we expect the intensity to drop to zero in regions where the waveguide is exactly half poled \cite{reitzig2021seeing}. The TPM image in Fig. \ref{fig:TPM} \textbf{(a)}, whose intensity at the center of the waveguide in the yellow box is shown in Fig. \ref{fig:TPM} \textbf{(b)}, reveals that the intensity in the unpoled waveguide is 3.4 (2.0) times higher than the minimum (peak) intensities in the unpoled section. This supports the fact that the poling depth is between 25 and 35 \% of the waveguide height. Note that to get a reliable ratio, we ensure that the TPM camera is not saturated at the highest SH power.

Second, a variation of the waveguide thickness and etch-depth along its length tends to broaden the SHG bandwidth and to decrease the peak conversion efficiency \cite{wang2018ultrahigh} accordingly. In fact, as the integral area of the SHG spectrum is preserved, we can estimate by how much the experimental peak efficiency is lowered compared to the uniform cross-section. To do so, we computed the SHG for a uniform waveguide and normalize it such that the area under its spectrum is the same as the experimental one. As shown in Fig. \ref{fig:SHG_spectrum_gradient_dimensions}, we obtain that the peak efficiency is reduced by a factor of 1.4. The above-mentioned analysis decreases the expected SHG conversion efficiency to the range 700 – 2100 \% W$^{-1}$cm$^{-2}$. 
Also, other qualitative arguments can explain a further reduction of the SHG conversion efficiency. Our electrodes are not continuous and have a duty cycle of 50 \% to ensure simultaneous detection of SHG by QPM and MPM. This results in regions between the electrodes where the poling depth is shallower than between the electrodes. Indeed, the TPM image in Fig. \ref{fig:TPM} shows some peaks of intensity in the poled section, corresponding to regions with shallower poling depth. Contrary to the QPM case, non-continuous poled regions, do not lead to interference effects of SH generated in different sections, but reduces however the effective length of the PPLN. 
Finally, the propagation losses and the collecting losses at the output might not be negligible as we assumed in our estimate. More precisely, we could not experimentally assess the propagation losses of the TE$_{01}$ mode at the SH wavelength in our waveguides and, as this mode has a high divergence, we might not collect all the SH light in our set-ups, under-estimating the SHG conversion efficiency. 

\section*{References}
\nocite{*}
\bibliography{MPM_Draft_supplementary_material_biblio}